\documentclass{nature}
\usepackage{amsmath, amsthm, amssymb, mathrsfs, graphicx, multibib, hyperref}
\usepackage{color,ulem}
\newcites{method}{Methods References}%
\usepackage{lineno}
\usepackage{longtable}
\newcommand{\sun}{$_\odot$}

\newcommand{\bapp}{\bar{\beta}_{\rm app}}

\definecolor{green}{rgb}{0.0, 0.5, 0.0}

\def\d{\mathrm{d}}
\newcommand{\thec}{\theta_{\rm c}}
\newcommand{\mr}{\mathrm}
\newcommand{\theLOS}{\theta_{\rm v}}

\newcommand{\epse}{\epsilon_{\rm e}}
\newcommand{\epsB}{\epsilon_{\rm B}}

\newcommand{\Eiso}{E_{\rm iso}}
\newcommand{\mc}{\mathcal}
\newcommand{\lara}[1]{\langle#1\rangle}

\title{Optical measurement of superluminal motion in the neutron-star merger GW170817}

\begin{document}

\maketitle


\noindent
\author{Kunal P. Mooley$^{1,2,*}$, Jay Anderson$^{3,*}$, Wenbin Lu$^{4,5,6,*}$}

\begin{affiliations}
\item Caltech, 1200 E California Blvd, MC 249-17, Pasadena, CA 91125, USA
\item National Radio Astronomy Observatory, Socorro, New Mexico, 87801, USA
\item Space Telescope Science Institute, 3700 San Martin Drive, Baltimore, MD 21218, USA
\item Department of Astrophysical Sciences, Princeton University, Princeton, NJ 08544, USA
\item Departments of Astronomy and Theoretical Astrophysics Center, UC Berkeley, Berkeley, CA 94720, USA
\item TAPIR, Walter Burke Institute for Theoretical Physics, Mail Code
  350-17, Caltech, Pasadena, CA 91125, USA
\end{affiliations}
\vspace{-1cm}
\noindent $^*$ These authors contributed equally to this work




\vspace{1cm}
\begin{abstract}
The afterglow of the binary neutron star merger GW170817\cite{abbott2017-gw170817} gave evidence for a structured relativistic jet\cite{mooley2018-wideoutflow,mooley2018-vlbi,ghirlanda2019,hajela2019,troja2020} and a link\cite{abbott2017-grb170817a,fong2017,mooley2018-vlbi} between such mergers and short gamma-ray bursts.
Superluminal motion, found using radio very long baseline interferometry\cite{mooley2018-vlbi} (VLBI), together with the afterglow light curve provided constraints on the viewing angle (14--28 degrees), the opening angle of the jet core ($<$5 degrees), and a modest limit on the initial Lorentz factor of the jet core ($\Gamma_{i}>4$).
Here we report on another superluminal motion measurement, at seven times the speed of light, leveraging Hubble Space Telescope precision astrometry and previous radio VLBI data of GW170817.
We thereby obtain a unique measurement of the Lorentz factor of the wing of the structured jet, as well as substantially improved constraints on the viewing angle (19--25 degrees) and the initial Lorentz factor of the jet core ($\Gamma_{i}>40$).

\end{abstract}



We carried out precision astrometric measurements of GW170817 using Hubble Space Telescope (HST) data obtained at mean epochs of 8 d and 159 d post-merger (each of the two measurements utilizes HST exposures taken over a span of several days, see Methods).
Our measurement at 8 d, when the optical emission was dominated by the thermal emission due to r-process nucleosynthesis (i.e. kilonova or macronova), indicates that the position of the neutron star merger is RA$=$13:09:48.06847(2), Dec.$=-$23:22:53.3906(2) (1$\sigma$ uncertainties  in the last digits are given in parentheses).
Our measurement at 159 d, when the optical emission was jet-dominated (non-thermal emission), indicates that the position of the afterglow was RA$=$13:09:48.06809(89), Dec.$=-$23:22:53.383(11).
While the precision of the former measurement rivals radio VLBI, the precision of the latter is coarse and would have benefited from a deeper HST observation at the peak of the afterglow light curve.
Positions of the optical source at both epochs are shown in Figure~\ref{fig:proper_motion}.

Astrometry tied to GAIA\cite{gaia-dr2-summary,gaia-edr3} enables us to analyze the optical and radio positions of GW170817 together.
Comparison of the 8 d HST measurement with the High Sensitivity Array (HSA) radio VLBI measurements\cite{mooley2018-vlbi} at 75 d and 230 d post-merger suggests offsets of $2.41\pm0.31\pm0.22$ mas and $5.07\pm0.33\pm0.22$ mas (1$\sigma$ uncertainties; statistical and systematic, respectively; see Methods), implying mean apparent speeds of $7.6\pm1.3$ and $5.2\pm0.5$ respectively, in units of speed of light. Here we have used the host galaxy distance of\cite{cantiello2018} $40.7\pm2.4$ Mpc (using the distance and associated uncertainty from ref.\cite{hjorth2017} does not change the apparent speeds to the specified significant digits).
With respect to the global VLBI radio position\cite{ghirlanda2019} at 206 d, the HST position is offset by $4.09\pm0.35\pm0.23$ mas, indicating motion at $4.7\pm0.6$ times the speed of light.
Offset positions of the optical and radio source along with the positional uncertainties are shown in Figure~\ref{fig:proper_motion}.
In comparison, the proper motion and the mean apparent speed measured with HSA\cite{mooley2018-vlbi} between 75 d and 230 d is $2.7\pm0.3$ mas and $4.1\pm0.5$ times the speed of light respectively.

For obtaining precise constraints on geometry and jet parameters, we consider the HST--HSA superluminal motion measurements. 
First, we use the point-source approximation and to estimate the true speed of the emitting material ($\beta$, in units of speed of light) and its angle with respect to the Earth line of sight ($\theta$) from the apparent speed $\beta_{\rm app}$.
In such a case we have $\beta_{\rm app} = \beta~\mathrm{sin}(\theta)/(1-\beta~\mathrm{cos}(\theta))$. 
Since $\beta$ is less than unity, the inclinations of the emitting regions at 75 d and 230 d are $<$18 degrees and $<$24 degrees (1$\sigma$ upper limits) respectively.
The material along the axis of the jet comes into view only around the time when the afterglow light curve starts declining steeply, occurring around\cite{mooley2018-strongjet,makhathini2020} $t_c\simeq175$ days post-merger, when the core has decelerated to a Lorentz factor of approximately the inverse viewing angle (i.e. $\Gamma_{175\d}\simeq1/\theta_v$, where $\theta_v$ is the viewing angle --- the angle between the jet axis and the Earth line of sight).
While we do not know the position of GW170817 around time $t_c$, we can constrain the mean apparent speed between 0 d--175 d to be larger than $5.2-0.5=4.7$ (1$\sigma$ lower limit) times the speed of light, leading to a conservative limit on the viewing angle of GW170817 of $<$24 degrees.

We now turn to estimating the orientation and Lorentz factor evolution of the jet wing.
The maximum value of the apparent speed, $\beta_{\rm app}=\Gamma\beta$, is obtained for $\beta=\mathrm{cos}(\theta)$ (i.e. for $\Gamma\gg1$ the maximum $\beta_{\rm app}=\Gamma$ occurs when $\Gamma=1/\theta$). Since we have measured the mean apparent speed $\bar{\beta}_{\rm app, 0\d-75\d}\simeq 7$ (but not the instantaneous apparent speed), the initial Lorentz factor of the material dominating the flux at 75d must have been $\Gamma_{i,75\d}\gtrsim 7$. Here we assume that the HST 8 d kilonova position denotes the position of the merger, and hence use the subscript ``0d--75d" for $\bar{\beta}_{\rm app}$.
We have denoted with the subscript ``$i$" the initial Lorentz factor (before deceleration) and with ``75d" the material that is dominating the afterglow emission at 75 days post-merger.
We can also estimate the instantaneous Lorentz factor $\Gamma_{75\d}$ of this jet wing material seen at 75 d in the observer's frame.
The mean apparent speed is given by, $\bar{\beta}_{\rm app,0\d-75\d} \simeq 8\theta_{75\d} \Gamma_{75\d}^2 / (4\Gamma_{75\d}^2 \theta_{75\d}^2 + 1)$ (see Methods).
For simplicity we assume 
that the region satisfying $\Gamma=1/\theta$ dominates the emission at any given time prior to the peak of the afterglow light curve.
Solving for the two parameters then we find $\Gamma_{75\d}\sim4.5$ and $\theta_{75\d}\sim13$ degrees. 
The HST--HSA measurement of superluminal motion therefore gives us a unique constraint on the Lorentz factor of the {\it wing} of the structured jet located approximately 13 degrees from the Earth line of sight.
This result
disfavors alternative models such as top-hat jet and refreshed shock\cite{gill2019,lamb2020} for the afterglow emission in GW170817.

We can use the above method to further estimate the viewing angle and the Lorentz factor of the jet core at 230 d, since the afterglow emission at this time should be dominated by the core (i.e. $\theta_{230\d}=\theta_v$).
In order to simultaneously satisfy (a) $\Gamma_{175\d}\simeq1/\theta_v$, (b) $\bar{\beta}_{\rm app,0\d-230\d} \simeq 8\theta_v \Gamma_{230\d}^2 / (4\Gamma_{230\d}^2 \theta_v^2 + 1)$  and (c) $\Gamma\propto t^{-3/8}$ (the Blandford-McKee evolution\cite{blandford76_dynamics}), the viewing angle is inferred to be $\theta_v\sim17$ degrees, and correspondingly $\Gamma_{230\d}\sim3.3$. In reality, the emission at a given time does not come from the region precisely satisfying\cite{beniamini2020} $\Gamma\theta=1$, so we calculate these viewing angles and Lorentz factors in a more detailed semi-analytical point-source model taking into account the likelihood distribution of $\Gamma\theta$ (described in Methods \S \ref{sec:point-source}). 
For the jet wing we obtain $\Gamma_{75\d} = 5.6^{+3.8}_{-1.7}$ and $\theta_{75\d} = 12.8^{+2.5}_{-2.5}$ degrees, and for the and jet core we find $\theta_v = \theta_{230\d} = 21.3^{+2.5}_{-2.3}$ degrees and $\Gamma_{230\d} = 4.7^{+3.1}_{-1.4}$ (1$\sigma$ uncertainties). 
These results are shown graphically in Figure~\ref{fig:speed_theta} (panel a). A schematic diagram showing the derived geometry of the wing and core of the structured jet in GW170817 can be found in Figure~\ref{fig:geometry}.

From the Lorentz factor of the emitting material at 230 d, we can also get a measurement of the ratio between the isotropic equivalent energy for the jet core $E_{\rm iso}$ and the density of the pre-shock medium $n$ as ${E_{\rm iso}/n} = 10^{55.8\pm 0.5}\rm\, erg\,cm^3$ (see Methods). It is not possible to obtain a robust constraint on $E_{\rm iso}/n$ based on the panchromatic afterglow light curves alone, because there is an additional free parameter $\epsilon_{\rm B}$ (the fraction of thermal energy in magnetic fields) that cannot be disentangled without measuring the characteristic synchrotron cooling frequency\cite{granot_kumar2003}.


For a robust verification of the above results, we used the relativistic hydrodynamic code $\mathtt{Jedi}$\cite{lu20_jet_dynamics} to carry out about a million independent simulations of an axisymmetric, structured jet interacting with the circum-stellar medium, including the effects of lateral expansion (see Methods).
We parameterize the angular dependencies of the kinetic energy and Lorentz factor structures of the jet using smoothed broken power-law functions.
The free parameters of the structured jet model are constrained based on the $\chi^2$-fits to the complete proper motion and afterglow lightcurve dataset of GW170817. 
The fits to the observational data are shown in Figure~\ref{fig:speed_theta} (panels b and c).
The modeling yields stringent constraints on the jet inclination angle $19\leq \theta_v \leq 25$ degrees, the Lorentz factor of the jet core $\Gamma_{i, c}>40$, the core opening angle $4\leq\theta_{c}\leq6$ degrees, all at 90\% confidence. This indicates that the results from the semi-analytic point-source model (which is applicable in the limits  $\theta_c\ll1\rm\, rad$ and $\theta_c\ll \theta_v$) are remarkably accurate and confirms that the HST positional measurement of GW170817 substantially improves the parameter constraints compared to those obtained from the radio VLBI positions alone. Our constraint is consistent with the initial Lorentz factors deduced for regular (on-axis) short-GRBs using compactness arguments and other techniques\cite{lithwick_sari2001,nakar2007,ghirlanda2018-Gamma,matsumoto2019}.
Also, the viewing angle is in agreement with the best-fit model found by ref\cite{mooley2018-vlbi}, but somewhat larger than that found by other studies\cite{ghirlanda2019,hotokezaka2019-h0} of GW170817 that jointly fit the afterglow light curve and VLBI proper motion. However, these latter studies (which find $\theta_v\simeq$14--17 degrees, 68\% confidence) are possibly biased\cite{hotokezaka2019-h0} to very low viewing angles due to the priors considered, and in any case agree with our hydrodynamical modeling result within the 90\% confidence interval.

Our study represents, to the best of our knowledge, the first proper motion constraint on the Lorentz factor of a GRB jet indicating ultra-relativistic ($\Gamma\gg10$) motion.
The limit $\Gamma_{i,c}>40$ cleanly separates GW170817 from Galactic systems, such as X-ray binaries having\cite{fender2006} $\Gamma\simeq1-7$ jets, as well as Active Galactic Nuclei and Tidal Disruption Events in which Lorentz factors up to $\Gamma\simeq40$ have been reported\cite{lister2013,mattila2018}.
While our limit $\Gamma_{i,c}>40$ implies low baryon-loading (ejecta mass $<10^{-4}$ M\sun) in GRBs like GW170817, the lower Lorentz factors measured in other systems might imply baryon-polluted jets.

\begin{figure}
\includegraphics[width=6.5in]{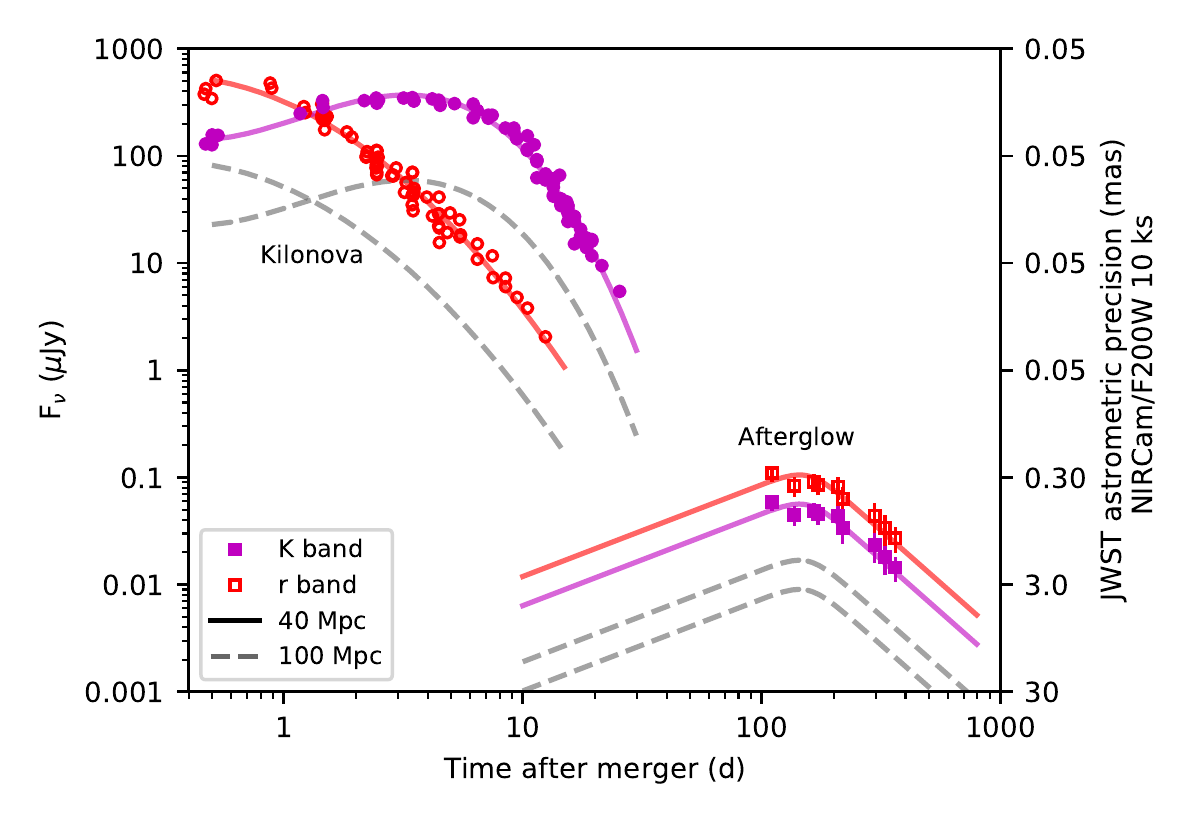}
\caption{{\bf Precision astrometry with the JWST.} The kilonova and afterglow light curves\cite{villar2017,makhathini2020} of GW170817 shown along with the astrometric precision expected using exposure time 10 ks with the JWST NIRCam (for the F200W filter; precision with the F070W filter will be a factor of 2.5 worse).
Thanks to its large collecting area and smaller pixel size, the astrometric precision of the JWST will be several times better than that of the HST for the same exposure time.
The dashed curves denote the kilonova and afterglow for a GW170817-like source at a distance of 100 Mpc.
Purple and red colors indicate K band and r band.
We assume a limiting astrometric precision of 0.05 mas.
}
\label{fig:jwst}
\end{figure}

We have demonstrated in this work that precision astrometry with space-based optical/infrared telescopes is an excellent means of measuring the proper motions of jets in neutron star mergers, and therefore also for constraining the geometries and Lorentz factors of such gravitational-wave sources.
The James Webb Space Telescope (JWST) should be able to perform astrometry much better than that with the HST, owing to the larger collecting area and smaller pixel size.
In Figure~\ref{fig:jwst} we show that, for a reasonable allocation of time, the JWST can achieve sub-milliarcsecond astrometric precision not only for the kilonova, but also for an afterglow like that of GW170817.
The combination of optical astrometry and radio VLBI measurements (with current observing facilities) may be even more powerful, and could deliver strong constraints on the viewing angles of neutron star mergers located as far away as 150 Mpc as long as they have favorable inclination angles and occur in relatively dense environments compared to GW170817.

\clearpage
\section*{References}
\bibliographystyle{naturemag}
\bibliography{refs}

\clearpage
\includegraphics[width=6.5in]{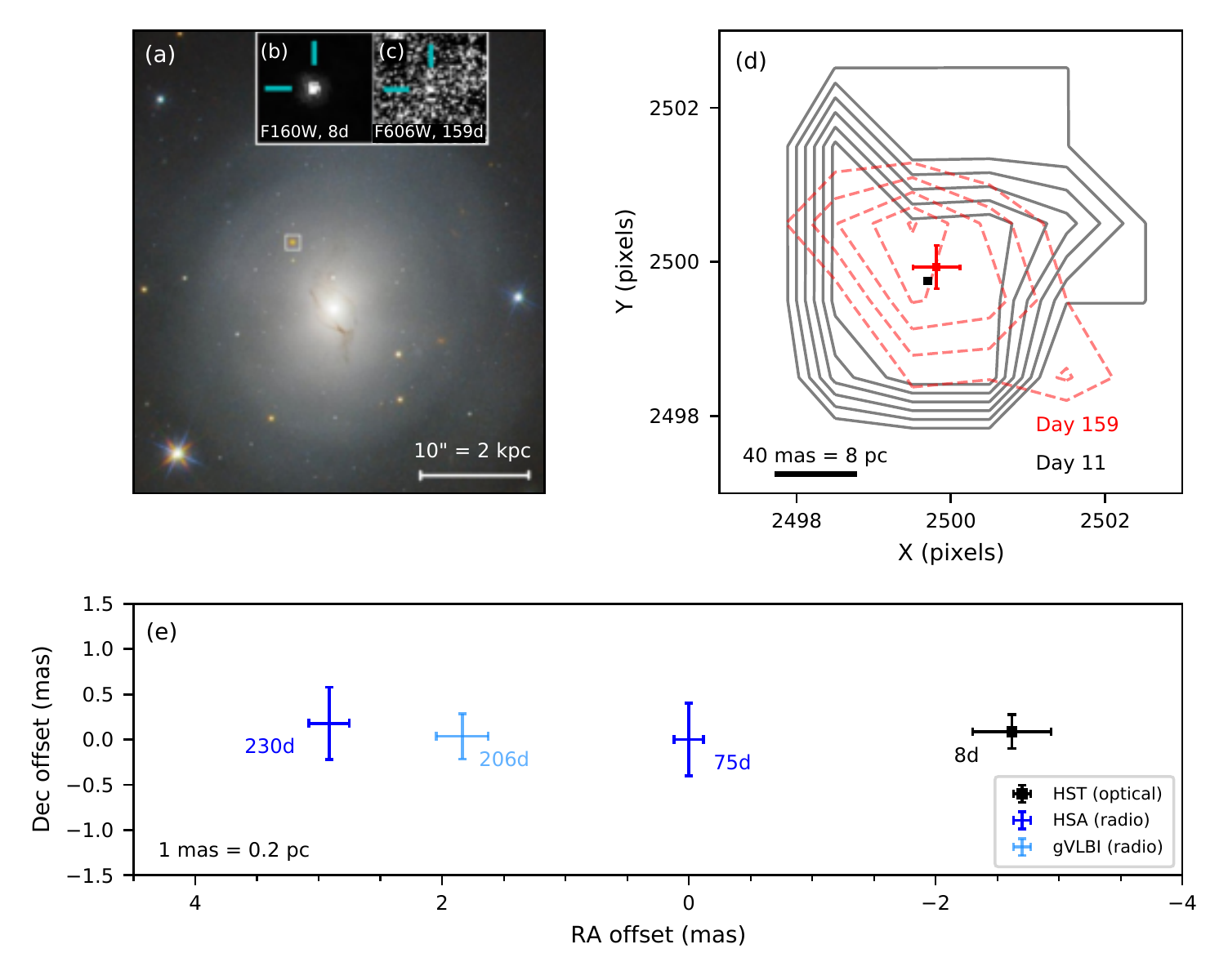}

\begin{figure}
\clearpage
\caption{{\bf Proper motion of GW170817.}
(a) Color composite HST image of the host galaxy NGC 4993. The white box denotes the GW170817 region zoomed in panels b,c. (b) The 2''$\times$2''~HST F160W stacked image of the kilonova (indicating the position of the merger) at mean epoch of 8 d post-merger. (c) The 2''$\times$2''~HST F606W stacked image of the afterglow at mean epoch of 159 d. (d) The positions of merger (black) and afterglow (red) on the GAIA pixel frame (see Methods). The contours are 24$\sigma$--49$\sigma$ and 2$\sigma$--7$\sigma$ in the HST stacked images from 8 d and 159 d respectively.
(e) The RA-Dec offset plot showing the position of GW170817 at 8 d post-merger, relative to the radio VLBI positions at 75 d and 230 d measured with the High Sensitivity Array\cite{mooley2018-vlbi} (HSA) and at 206 d with a 32-telescope global VLBI (gVLBI) array\cite{ghirlanda2019}. The 75 d VLBI measurement has offsets (0,0) as per the convention of ref\cite{mooley2018-vlbi}. 
All the radio VLBI positions have been transformed into the ICRF3 frame (see Methods).
In all panels, 1$\sigma$ statistical errorbars are shown (systematic uncertainties not shown). 
The proper motion measured between 8 d--75 d, 8 d--206 d, and 8 d--230 d is $2.41\pm0.38$ mas, $4.09\pm0.42$ mas and $5.07\pm0.40$ mas, implying mean apparent speeds of $7.6\pm1.3$, $4.7\pm0.6$ and $5.2\pm0.5$ in units of speed of light, respectively (1$\sigma$ statistical and systematic uncertainties added in quadrature). 
The 159 d HST measurement has a coarse precision and is not plotted in this panel.}
\label{fig:proper_motion}
\end{figure}

\clearpage
\includegraphics[height=2.9in,viewport=5 5 350 165,clip]{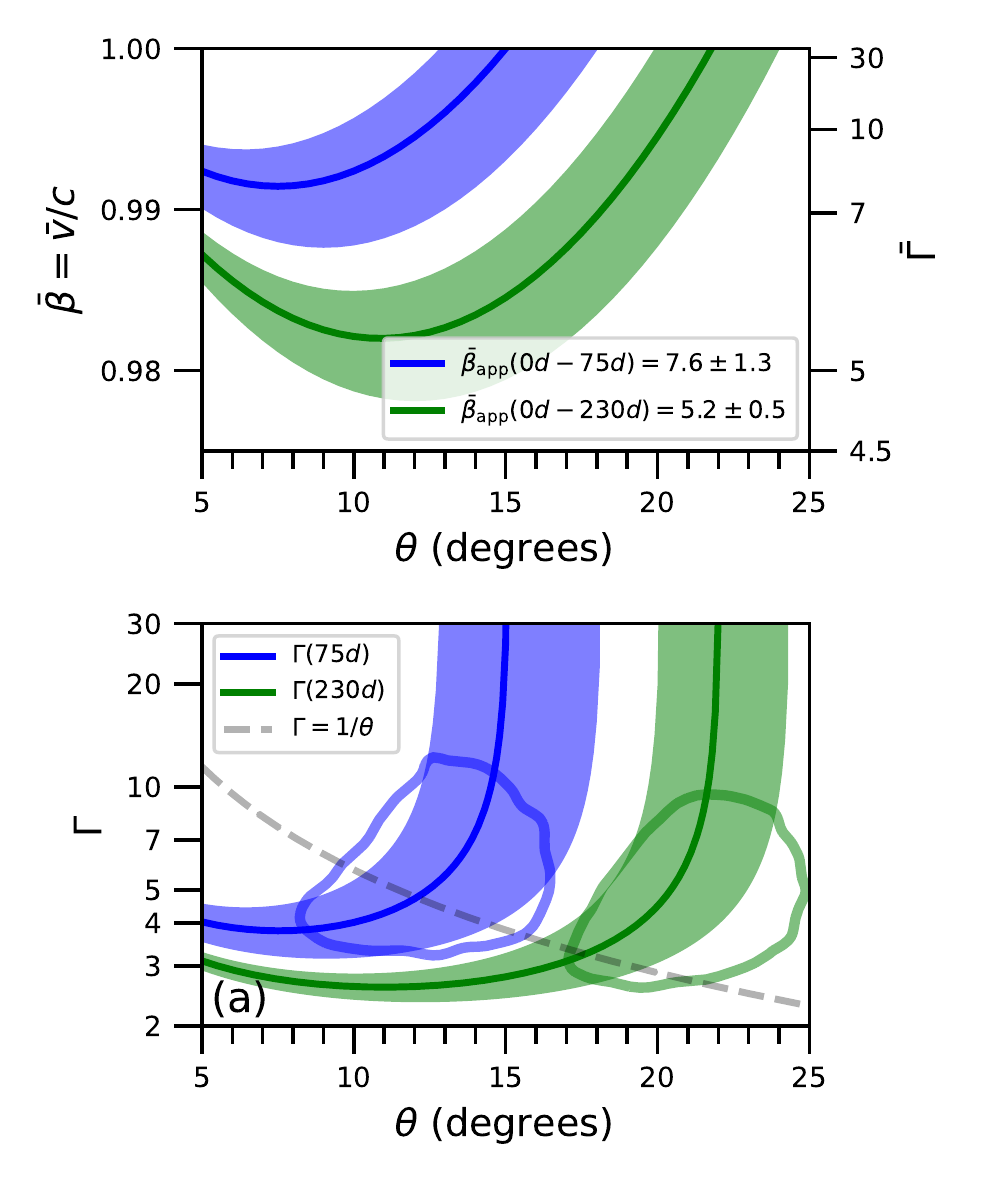}\\
\includegraphics[width=7in,]{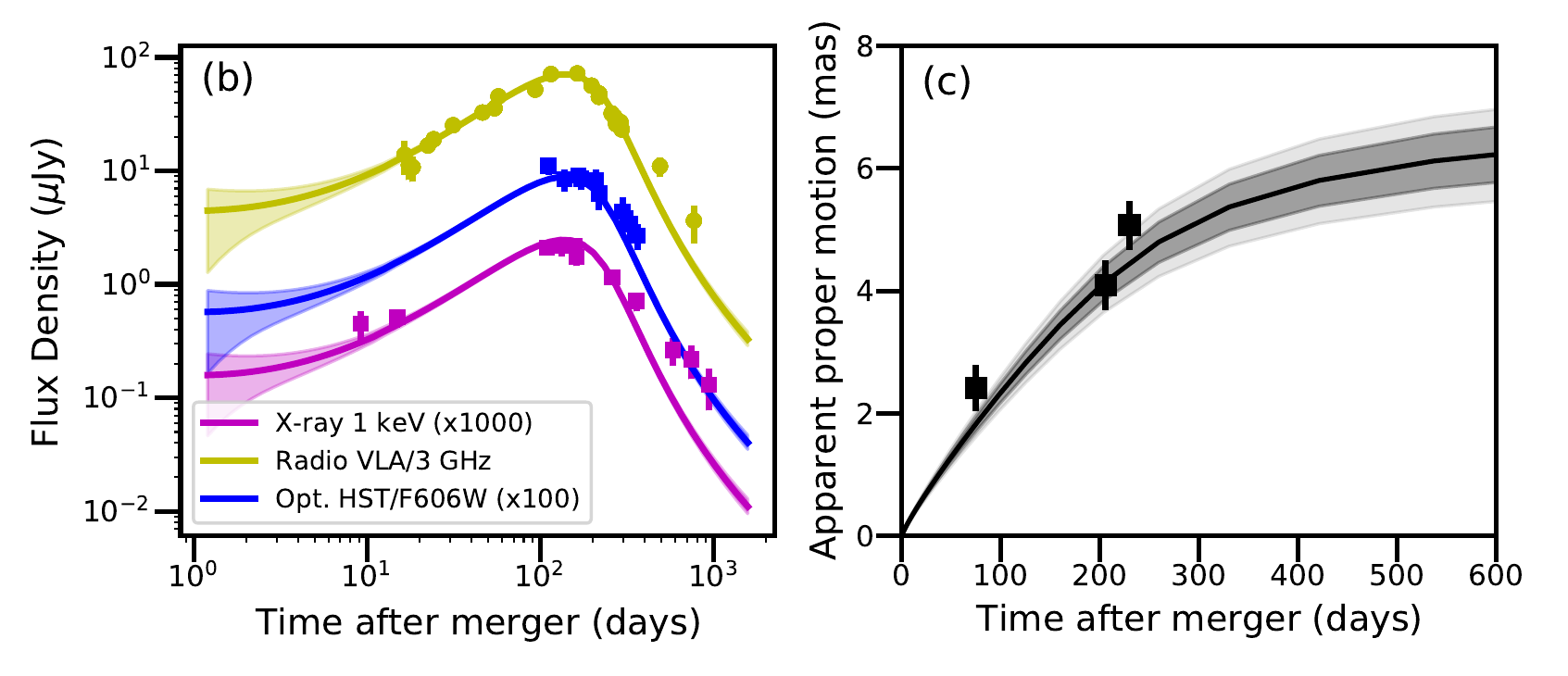}

\begin{figure}
\clearpage
\caption{{\bf Parameter estimations using the semi-analytical point source and hydrodynamical models.} 
(a) The Lorentz factors ($\Gamma_{75\d}$ and $\Gamma_{230\d}$, shown in blue and green) of the material dominating the afterglow emission at 75 d (jet wing) and 230 d (jet core) post-merger, as functions of their respective angles from the Earth line of sight ($\theta_{75\d}$ and $\theta_{230\d}=\theta_v$; see also Figure~\ref{fig:geometry}), obtained from the point-source model.
Angle $\theta_{230\d}$ corresponds to the material lying along the jet axis (jet core) as inferred from the afterglow light curve\cite{mooley2018-strongjet,hajela2019,troja2020}.
The blue and green contours (68\% confidence) denote the parameter space inferred from the semi-analytical model: $\Gamma_{75\d} = 5.6^{+3.8}_{-1.7}$, $\theta_{75\d} = 12.8^{+2.5}_{-2.5}$ degrees and $\Gamma_{230\d} = 4.7^{+3.1}_{-1.4}$, $\theta_v = 21.3^{+2.5}_{-2.3}$ degrees (see Methods). 
The dashed grey line denotes the approximation $\Gamma=1/\theta$ (just for reference) for the structured jet material dominating the afterglow emission at any given time.
(b), (c) Fits to afterglow light curve and proper motion data using the hydrodynamical simulations described in the Main Text and Methods. Only a subset of the full light curve data (total 104 data points including upper limits; see Methods), used in the model fitting, are shown in panel (b). All error bars are 1$\sigma$. The solid lines represent the median and the shaded areas represent the 68\% confidence intervals. The late-time discrepancy (three X-ray data points and two radio data points) between the jet afterglow model and the light curve data, seen in panel (b), has been noted by previous studies\cite{hajela22_excess, troja22_excess, balasubramanian21_radio_late_time} and interpreted as a slower-spreading jet or a new afterglow component.
Since the discrepancy only exists for 5\% of the full light curve dataset, and additionally only for the data taken well beyond the peak of the afterglow light curve, this issue does not significantly affect the jet parameter estimated from our hydrodynamical analysis.}
\label{fig:speed_theta}
\end{figure}

\clearpage
\begin{center}
\includegraphics[width=4.5in]{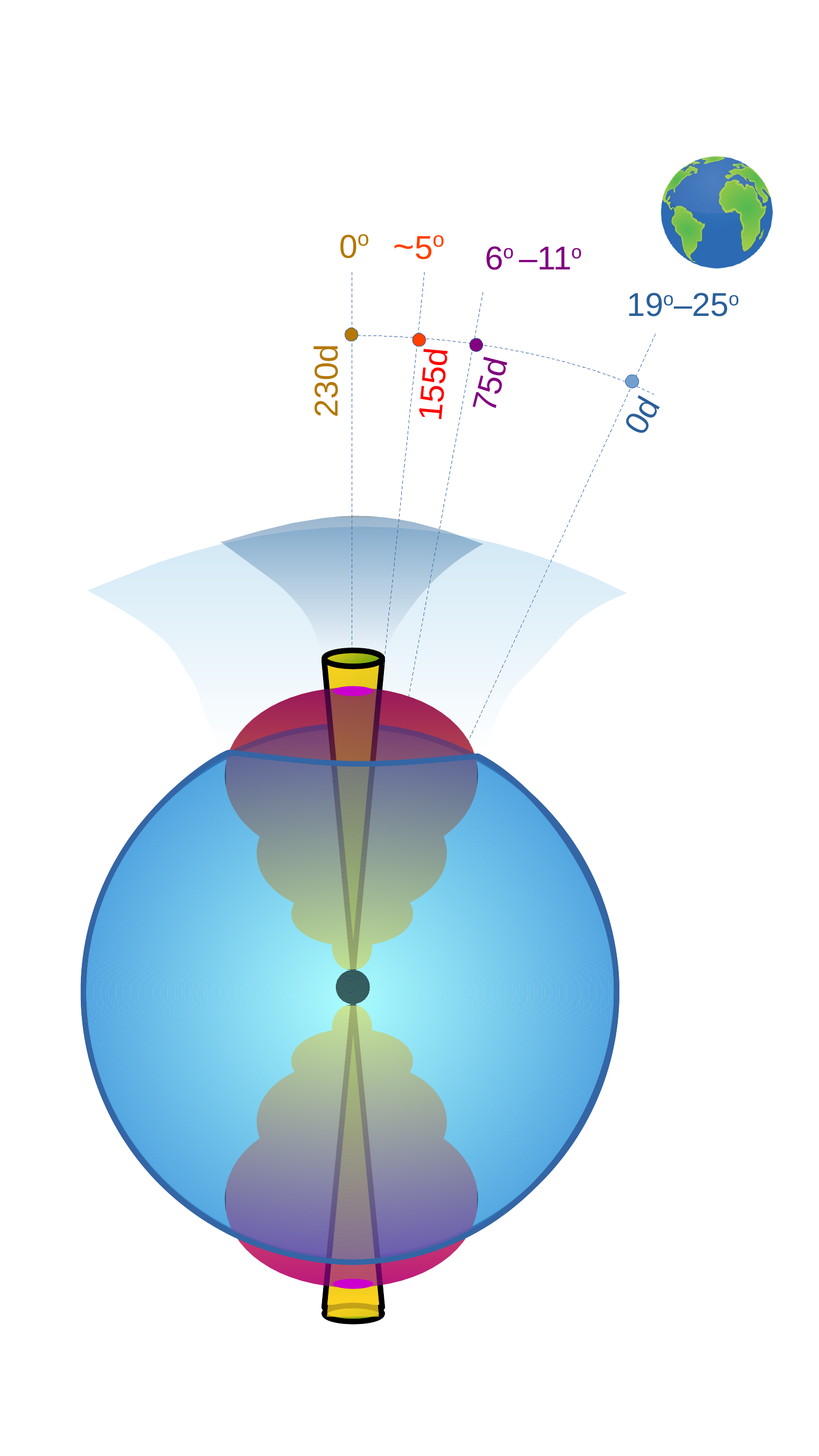}
\end{center}

\clearpage
\begin{figure}
\caption{{\bf Schematic of the geometric parameters derived for GW170817.} 
The jet core (yellow) and the surrounding cocoon or ``wing'' (red) produced through interaction with the dynamical ejecta (blue) are shown. The polar angles of the material dominating the afterglow emission at various observing epochs, 75 d, 155 d and 230 d post-merger, from the jet axis, are found to be 6--11 degrees (68\% confidence, based on semi-analytic point-source model), $\sim$5 degrees (based on hydrodynamic simulations), and 0 degrees (based on the afterglow light curve evolution\cite{mooley2018-strongjet,hajela2019,troja2020}). The angle between our line of sight and the jet axis is constrained to be 19--25 degrees (90\% confidence), based on our hydrodynamical simulations.
}
\label{fig:geometry}
\end{figure}

\clearpage

\clearpage
\begin{methods}

\section{Precision astrometry with the Hubble Space Telescope (HST)}

We used images from the HST Wide Field Camera 3 (WFC3) and Advanced Camera for Surveys (ACS) collected using filters F160W (IR channel) and F606W (UVIS channel), where GW170817 was sufficiently bright and was observed at multiple epochs. 
A summary of all the archival data used for the precision astrometry is given in Extended Data Table~\ref{tab:obs}.
We also inspected F110W exposures, but found that they are extremely undersampled and hence did not include them in the final analysis.

%

\subsection{F160W analysis.}
We analyzed only the pipeline-product \_flt images, which are flat-fielded slope images from the up-the-ramp sampling of the WFC3/IR detector.  
Since the gradient of the host galaxy NGC 4993 can affect astrometry of point sources superposed on it, we removed the galaxy profile by modeling the light distribution in each 1014$\times$1014 image with an array of 127$\times$127 points, each of which in turn represents the sigma-clipped average value of the image over an 8$\times$8-pixel region.  
We iteratively solved for the values of the representative grid by subtracting the current grid-model (interpolated with a bicubic spline) and examining the residuals within a 23$\times$23 box about each grid-point.  
In this way, we converged upon a smooth version of the background.  
Subtracting this background from the images allows us to measure the point sources (the reference stars and GW170817) without bias from the gradient of the galaxy.
We then brought all exposures to a common astrometric frame using the following steps.

First, for facilitating comparison with the radio VLBI astrometric data\citemethod{mooley2018-vlbi,ghirlanda2019}, we defined a pixelized GAIA astrometric reference frame at the 2017.65560 epoch. 
This frame is centered on the nominal GW170817 location, (RA,Dec) = (13:09:48.06900,-23:22:53.4000) $=$ (197.45028750 deg,-23.38150000 deg), has a tangent-plane pixel scale of 40 mas per pixel, and has the above nominal GW170817 location at pixel coordinate (2500.00,2500.00). 
The 40 mas per pixel was chosen because it corresponds to the WFC3/UVIS scale.
Second, we solved for and applied the HST distortion correction (described in the next section in detail) for each exposure.  

Third, we selected good GAIA DR2/EDR3\citemethod{gaia-dr2-summary,gaia-edr3} reference stars that were well measured in all of the seven F160W exposures (see Extended Data Table~\ref{tab:obs}). 
There are 32 GAIA stars that are within the WFC3/IR frame.  
The positions of these stars in the pixelized reference frame and their GAIA positional error are shown in Extended Data Figure~\ref{fig:gaia_sel}.  We vetted all these stars and shortlisted ``good'' stars that satisfy the following criteria: 1) low quoted GAIA positional errors ($<$0.6 mas)
%
%
2) not too close to the host galaxy nucleus ($>$12 arcseconds from the nucleus of NGC 4993), 3) lies within the CCD chip, and 4) not in the vicinity of any bad pixel.  
This yielded 7 good stars, out of which one was appeared to be a visual binary in the HST images. 
We therefore shortlisted 6 GAIA reference stars.
Fourth, the (X, Y) positions and associated uncertainties of these stars were calculated in the GAIA pixelized reference frame using the RA, Dec and proper motion from the GAIA EDR3 catalog\citemethod{gaia-edr3-catalog} and standard propagation of uncertainties.
The coordinates and other details for the 6 reference stars are given in Extended Data Table~\ref{tab:f160w_ref_stars}.
 
Next, the transformations from the HST images into the GAIA frame were effected by taking the positions of the 6 good stars in the pixelized GAIA frame and the distortion-corrected positions for the same stars in each of the four HST frames.
All positional measurements in the HST images were made using the point-spread function (PSF) fitting technique as detailed in ref\citemethod{anderson2016}, and are 
given in Extended Data Table~\ref{tab:transformation}.

From previous investigations of HST data we have found that a full 6-parameter linear transformation is needed to go from HST coordinates to GAIA.   
This is because HST ``breathes'' during its orbit around the Earth, and there is no available model to account for this.   
Breathing can introduce both scale changes and some off-axis linear terms.  Velocity aberration also introduces a scale change.  
A general linear transformation addresses both these issues implicitly.  
Such a transformation has the form: 

$
\begin{bmatrix}
X_{\rm GAIA} - X_{\rm GAIA,0}\\
Y_{\rm GAIA} - Y_{\rm GAIA,0}
\end{bmatrix}
=
\begin{bmatrix}
A & B\\
C & D
\end{bmatrix}
\begin{bmatrix}
X_{\rm COR} - X_{\rm COR,0}\\
Y_{\rm COR} - Y_{\rm COR,0}\\
\end{bmatrix}
$
\newline\newline
\noindent where (X$_{\rm GAIA}$, Y$_{\rm GAIA}$) are the transformed positions in the GAIA  pixelized frame, [A B, C D] is the transformation matrix and (X$_{\rm COR}$, Y$_{\rm COR}$) are the distortion-corrected positions in HST images.

%

Since one of the offsets (X$_{\rm COR,0}$, Y$_{\rm COR,0}$) or (X$_{\rm GAIA,0}$, Y$_{\rm GAIA,0}$) is arbitrary, this equation actually has 6 free parameters.  
We solve for the 6 parameters using weighted least squares technique.  
We need a minimum of 3 pairs of positions, so 3 stars for which we have a position in both frames will specify the transformation.  
Here, we have an over-constrained problem, since we have for each exposure 6 stars with positions both in the GAIA frame and the distortion-corrected HST frame.
%
%
%
%

Thus, since we have more constraints than free parameters, we inspected the residuals of the transformation to get a sense of how well our HST-GAIA associations agree with each other and to see how much we can trust the transformation.   
To this view, we back-calculated (X$^{\prime}_{\rm GAIA}$,Y$^{\prime}_{\rm GAIA}$) from the input (X$_{\rm COR}$, Y$_{\rm COR}$) positions and the transformation matrix and then compared the star positions with their original input GAIA positions.
We thus found the HST-GAIA residuals to be $<$0.3 mas, consistent with the GAIA positional errors, indicating that the transformation is robust and it is not introducing significant uncertainties in addition to the GAIA errors.

\subsection{F606W analysis.}
There are several HST observations of the kilonova in F606W, but many of them were taken with subarrays and there are very few GAIA stars available in the subarray field-of-view to allow an absolute astrometrization of the frame. 
Further, stars of different brightness are affected differently by charge-transfer efficiency (CTE) losses, and although there exists a CTE correction it is not perfect.
There are uncertainties in the CTE correction especially for images with relatively high backgrounds, like in the vicinity of NGC 4993. 

Nevertheless, we attempted precision astrometry on the late-time (afterglow) observations, which were undertaken primarily in the F606W filter (no subarrays were used; see Extended Data Table~\ref{tab:obs}). 
We examined these exposures in an effort to measure a proper motion between the HST kilonova position (from F160W, see above) and HST afterglow position (from F606W). 
To account, however imperfectly, for CTE we used the pipeline-product \_flc images, with the galaxy profile subtracted as described above for the WFC3/IR images.  
We then corrected the measured positions for distortion\citemethod{bellini2011}.
Since the GAIA reference stars used for the F160W analysis were almost all saturated in the deep F606W exposures, we transformed the F606W HST images into the GAIA pixelized frame by using the positions of $\sim$15 (depending on the field overlap in different images) of the medium-brightness stars in the WFC3/IR source catalog (these were too faint to be found in GAIA).
We note that the first afterglow observation was carried out in December 2017, when GW170817 came out of HST's solar-avoidance zone, and soon after there was a steep decline in the afterglow light curve\citemethod{dobie2018,mooley2018-strongjet,alexander2018,troja2018}.
Unfortunately, the December 2017 observation is not deep enough and the signal-to-noise ratio (SNR) in that HST image (and also in the subsequent F606W observations) is low\citemethod{lamb2019,fong2019,piro2019,makhathini2020} (SNR$\ll$10). 
We therefore coadded WFC3/UVIS and ACS/WFC data obtained between December 2017 and March 2018 in order to increase the SNR and measure a precise position of GW170817.

\section{WFC3/IR Distortion Correction}
The distortion correction places the stars at their true locations (X$_{\rm COR}$, Y$_{\rm COR}$) relative to the central pixel of the detector.  
The HST correction is typically a 3--4 order polynomial and usually has a fine-scale look-up-table component, which can depend slightly on the filter.  
One of us (J.A.) developed a distortion solution for WFC3/IR in 2010, based on commissioning observations of the center of Omega Centauri, and has been using it for scientific reductions since then.  
To both evaluate and improve the solution, we downloaded more than 100 F160W exposures from the archive of the cluster core taken between 2009 and 2020, at a variety of orientations and offsets.  
Since the stars have considerable internal motions at the center of Omega Centauri (0.01 WFC3/IR pixel per year), we could not compare all the images with each other, so we compared each image against the other images that were taken within 1.5 years in time.  
This gave us over 3000 image-to-image comparisons, and we distilled the many star residuals into a single plot.
We found small residuals (0.005 pixel) and using these developed an improved distortion correction.
These residuals in the X and Y positions before and after the improved distortion correction are shown in Extended Data Figure~\ref{fig:distortion_correction}.   
In general, the residuals went down by a factor of two (root-mean-square, RMS), so that the new residuals are within 0.002 pixel per coordinate (i.e. within 0.08 mas).
We find that the distortion correction does not change significantly over time for the WFC3/IR detector.

\section{HST source position measurements and error estimation}
\label{sec:method:positions_error}

Above we have described how the HST images were transformed into the GAIA pixelized frame.
Here we describe the GW170817 positional measurements in these images (which are aligned to the GAIA frame) and the uncertainties associated with the positions. There are the following uncertainties in our analysis: 1) positional uncertainties of stars in HST frame, 2) uncertainty associated with the HST to GAIA/ICRS transformation, and 3) uncertainty in the measured optical position of GW170817.
We investigate these sources of uncertainties below.
 
\subsection{GW170817 positional measurements, HST errors for GW170817 and reference stars.}

For each of the seven F160W exposures (see Extended Data Table~\ref{tab:obs}), we measured the optical positions of GW170817 and field stars using the PSF fitting procedure described in ref\citemethod{anderson2016} (as done for the GAIA reference stars, described above).
The GW170817 positions are given in Extended Data Table~\ref{tab:transformation}.
For the positional uncertainty, we take the empirical uncertainty as the standard deviation of several field stars (located within the CCD chip; including GW170817) in the exposures, and disregard the statistical uncertainty associated with the PSF fits. 
This has the advantage of incorporating all uncertainties associated with the transformation, distortion correction, and other unknown contributors in the HST data, into the error estimate.
For the exposures obtained on 22 August 2017 and 27 August 2017 we find that the empirical uncertainties in (X, Y) coordinates are (0.022 pix, 0.009 pix) and (0.017 pix, 0.020 pix) respectively.
The relative positional uncertainties at these two epochs therefore roughly scale inversely as the detection SNR of GW170817 ($\sim$370 and $\sim$270 in each exposure of the 22 August and 27 August respectively; see Extended Data Table~\ref{tab:obs}) and, converting back to the native pixel scale for the WFC3/IR detector (120 mas/pixel), imply an achieved precision of $\simeq$(2\,CCD\,pixels)/SNR in the positional measurements, consistent with expectations for HST data.

Finally, we combine the GW170817 positions from the seven F160W exposures by taking the weighted mean and the associated uncertainty, to obtain the final position of (X, Y) $=$ (2500.182$\pm$0.002, 2500.235$\pm$0.001) (mean$\pm$error in X, Y coordinates), 
which implies RA$=$13:09:48.068473(5), Dec$=-$23:22:53.39059(4) or equivalently, RA$=$197.45028530(2) deg, Dec$=$-23.38149738(1) deg at a mean epoch of 8 d post-merger.
The positions of GW170817 for each of the F160W exposures in the pixelized GAIA frame, and the final combined position, are shown in Extended Data Figure~\ref{fig:consolidated_positions}.
This analysis includes the errors in the HST positions of the reference stars, 
but does not include the GAIA errors in the reference stars used for the frame transformation.
We investigate this point in the following subsection.

For the F606W filter data for the afterglow of GW170817, we prepared an image stack using observations taken between December 2017 and March 2018 (as mentioned above; mean epoch 159 d) since the afterglow was faint.
We measured the position using JMFIT in AIPS\cite{greisen2003} taking the Gaussian axial parameters obtained from the fitting of a nearby relatively bright star (coordinates 13:09:48.069 -23:22:55.81, located 2.5 arcsec to the South of GW170817). 
We fitted for the peak and position of GW170817 and found the best-fit position to be (X, Y) $=$ (2500.31$\pm$0.30, 2500.43$\pm$0.28), i.e. RA$=$13:09:48.06809(88), Dec$=-$23:22:53.383(11).
Since the precision on the afterglow position is low (12~mas), it is not useful for calculating proper motion and we do not further explore the systematic contributions to the F606W errors.
We note that a deep HST observation in December 2017 could have substantially improved the precision, $\mathcal{O}$(1mas), on the afterglow position, thereby facilitating an HST-only measurement of superluminal motion (without relying on radio VLBI positional measurements).

\subsection{GAIA errors.}

To understand how the positional errors of the GAIA reference stars (used for computing the frame transformation matrix for the F160W filter data) within the GAIA catalog might affect our analysis, we did a Monte Carlo-type (MC-type) simulation. 
We took each GAIA star's position and added a Gaussian deviate of its formal GAIA error to its X and Y position. 
We then recomputed the transformations and the positions of GW170817 for each exposure, then for each MC sample calculated a weighted-mean position for GW170817 using the empirical positional uncertainties described above.  
Taking all MC samples together we found the standard deviation of X and Y positions to be 0.007 pixel and 0.005 pixel respectively, corresponding to 0.31 mas in RA and 0.18 mas in Dec. 
Thus, we find that the uncertainty in the GW170817 position is dominated by the GAIA errors. 

\subsection{Other possible sources of error.}

From Extended Data Figure~\ref{fig:distortion_correction} we see that the distortion-correction residuals lie within 0.002 pixel per coordinate (i.e. within 0.08 mas; RMS).
We have also shown earlier that the HST-to-GAIA frame transformation residuals are consistent with the expected GAIA position errors, which implies that there is no significant transformation error. 
Nevertheless, any residual error in the distortion solution, or the frame transformation, or from an unknown origin in the HST data, should be included in the error analysis presented above (since we consider empirical uncertainty in the position of GW170817).
Therefore, the error in the mean GW170817 position (0.07 mas in RA and 0.04 mas in Dec) together with the error resulting from the GAIA reference star positional uncertainties (0.31 mas in RA and 0.18 mas in Dec) should adequately characterize the error.
We add these two contributions in quadrature to calculate the formal error in the HST position of GW170817, 0.32 mas in RA and 0.19 mas in Dec, and thus the final positional measurement at 8 d as RA$=$13:09:48.06847(2), Dec$=-$23:22:53.3906(2) or equivalently, RA$=$197.45028530(8) deg, Dec$=-$23.38149738(5) deg.
This position of GW170817 together with its positions at other epochs, considered for proper motion measurement, is given in Extended Data Table~\ref{tab:coordinates}.

 
%
 %

\section{Correction to the radio VLBI positions of GW170817 and associated errors}
Mooley et al. 2018\cite{mooley2018-vlbi} (hereafter MDG18) used J1258-2219 (2.7 degrees away from GW170817) and J1312-2350 (0.8 degrees away) as the primary and secondary phase referencing sources.
Hence, the MDG18 positions of GW170817 are in a J1312-2350-based coordinate frame tied to the position of J1258-2219.
We therefore seek a precise position of J1312-2350 in the GAIA or ICRF3 frame to find the correct positions of GW170817 at 75 d and 230 d for comparison with the HST 8 d position calculated in the previous section. 

First, we note that MDG18 used the position 12:58:54.4787760~ $-$22:19:31.125540 for J1258-2219 from the RFC2015a catalogue\citemethod{rfc_url} (which was, at the time, standard with the VLBI SCHED 11.4 program\citemethod{sched_url}), but we found a revised ICRF3 position (ICRF3 source catalogs from the Goddard Space Flight Center VLBI group\citemethod{icrf3_gsfc_url}, generated 2021-APR-05), 12:58:54.4787818(37) $-$22:19:31.12504(10).
Therefore, the positions of J1258-2219 and J1312-2350 need to be corrected;  $0.08\pm0.05$ mas and $0.50\pm0.10$ mas should be added to the RA and Dec. respectively to bring the source coordinates to the ICRF3 frame.

Second, the VLBI position of J1312-2350, determined based on phase referencing J1258-2219, from MDG18 (measured through Gaussian fitting of the source with AIPS/JMFIT) is 13:12:48.7580627(1) $-$23:50:46.95309(3) (Adam Deller, private communication), so the corrected ICRF3 position is 13:12:48.758068(3) $-$23:50:46.9526(1).
However, there is a relatively large systematic uncertainty associated with this position.
Since J1258-2219 and J1312-2350 are separated by 3.5 degrees we estimate that the systematic uncertainty, arising from phase referencing and ionospheric contribution, in this position should be about 0.2 mas in RA and 0.6 mas in Dec (ref\citemethod{pradel2006} and Adam Deller, private communication; note that $\sim$5 full-track VLBI observations were carried out by MDG18).
This uncertainty was not relevant for the proper motion measurement made by MDG18 since both their positional measurements of GW170817 were referenced directly to J1312-2350. In this work, however, we want to bring all positions to the GAIA or ICRF3 reference frames so we need to take these uncertainties into account.

Third, we find that there are two additional positional measurements available for J1312-2350.
One from the GAIA-EDR3 catalog, 13:12:48.758072(9) $-$23:50:46.9530(1), and the other from absolute astrometry in the radio\citemethod{rfc2021b_url}, 13:12:48.758111(37) $-$23:50:46.9532(14).
The position corrected to ICRF3 from the AIPS/JMFIT measurement (described above) agrees with the GAIA-EDR3 position to within 0.05$\pm$0.13$\pm$0.20 mas in RA and 0.39$\pm$0.14$\pm$0.60 mas in Dec (1$\sigma$ uncertainties; statistical and systematic respectively), and with the RFC2021b position to within 0.58$\pm$0.51$\pm$0.20 mas in RA and 0.59$\pm$1.35$\pm$0.60 mas in Dec.
The excellent agreement between all these positions (ICRF3 position corrected from MDG18, GAIA-EDR3 position, and ICRF3 position from RFC2021b) of J1312-2350 within 1$\sigma$ uncertainties suggests that we can use the three measurements to obtain a precise (weighted mean) position of this radio calibrator source. 
For the GAIA-EDR3 position, however, we will have to first take into account the radio-optical position offset due to different emitting regions at the two observing frequencies (i.e. the core-shift effect\citemethod{kovalev2017,petrov2017,petrov2019}).
The median offset between ICRF3 and GAIA sources is found to be\citemethod{charlot2020-icrf3} 0.58 mas, so we add 0.58/$\sqrt{2}$ mas in quadrature with the RA and Dec uncertainties of the GAIA-EDR3 position and then find the weighted mean of all three positions of J1312-2350 as 13:12:48.758073(12) $-$23:50:46.9529(3).

Fourth, a comparison between this weighted mean position of J1312-2350 and the AIPS/JMFIT position from MDG18 implies that 1) the MDG18 radio source positions of J1312-2350 and GW170817 need an additive correction of $0.14\pm0.18$ mas in RA and $0.21\pm0.34$ mas in Dec, and 2) the systematic uncertainties 0.18 mas in RA and 0.34 mas in Dec should be propagated to the uncertainties in the radio VLBI positions of GW170817 reported by MDG18.
It should be noted here that this uncertainty can be reduced to $\lesssim0.1$ mas (Adam Deller, private communication) in each coordinate with a dedicated radio astrometric observation of J1312-2350, where the calibrator is phase-referenced to a few nearby ICRF3 sources.
However, this uncertainty, although significant, does not dominate the uncertainties on our HST-VLBI proper motion measurements (as we show below), and hence we proceed with carrying these uncertainties through standard error propagation. 

We can use the MDG18 positions of GW170817, 13:09:48.068638(8) $-$23:22:53.3909(4) at 75 d and 13:09:48.068831(11) $-$23:22:53.3907(4) at 230 d, to compute its ICRF3 positions, 
13:09:48.068648(8) $-$23:22:53.3907(4) and 13:09:48.068841(11) $-$23:22:53.3905(4) at the two respective epochs (numbers in brackets indicate statistical-only uncertainties in the last digits of the RA and Dec).
These positions are shown in Figure~\ref{fig:proper_motion}.

Since ref\citemethod{ghirlanda2019} also used J1312-2350 as a phase calibrator, we can similarly compute the ICRF3 position of GW170817 at 206 d post-merger, 13:09:48.068770(14) $-$23:22:53.3906(3).

The final radio and optical positions of GW170817, in the GAIA or ICRF3 reference frame, together with the associated uncertainties are given in Extended Data Table~\ref{tab:coordinates}.



\section{Statistical and systematic uncertainties in the HST-VLBI proper motion measurements}
We consider the following possible contributions to the uncertainties in the proper motion measurements.

\subsection{Match between the GAIA and VLBI coordinate systems.}

The radio VLBI/ICRF3 reference frame has been found to agree with the GAIA-DR2 frame (called the GAIA-CRF2) to within\citemethod{mignard2018,liu2018,liu2020,charlot2020-icrf3} $\sim$30 $\mu$as or better for each axis, so we can neglect this contribution to the error budget.

\subsection{VLBI uncertainties.}

In the previous section we considered all uncertainties associated with the phase calibrator sources and arrived at the ICRF3 positions of GW170817 at 75d, 206 d and 230d (and corresponding statistical uncertainties). 
We additionally need to consider the systematic uncertainty arising from the phase referencing between J1312-2350 and GW170817, which MDG18 quoted as 0.15 mas in RA and 0.5 mas in Dec at each epoch. 

While calculating the proper motions of GW170817 between our HST 8 d position and the two VLBI positions, at 75 d and 230 d (and similarly for 206 d), we used standard propagation of uncertainty to calculate the total statistical and systematic uncertainties (the only systematic contributions are from VLBI). 
Finally, for each proper motion measurement, 8 d--75 d and 8 d--230 d (and similarly for 8 d--206 d), we added these statistical and systematic uncertainties in quadrature to get the total uncertainty on the superluminal motion, as quoted in the Main Text.

We note that the Gaussian uncertainties 0.18 mas and 0.34 mas in RA and Dec respectively on the radio VLBI measurements at 75 d, 206 d and 230 d, described in the previous, are correlated between the three radio measurements (since the same source J1312-2350 was used for phase referencing).
This correlation should, in principle, be taken into account during theoretical modeling of the proper motion data.
However, since the contribution of this correlated term to the total error budget in the proper motion and superluminal motion measurements (see Main Text and Extended Data Table~\ref{tab:coordinates}) is relatively small, $\sim$15--20\%, we simply assume that all the uncertainty terms are uncorrelated during the modeling (described below).

\subsection{Radio and optical positions of the host galaxy nucleus.}
The excellent agreement between the ICRF3 and GAIA-EDR3 positions of J1312-2350 gives shows that the offset between the radio VLBI images reported by MDG18 and the GAIA frame, to which our HST images are aligned, is negligible.
One additional check of the consistency between these two coordinate systems is the position of the nucleus of the host galaxy NGC 4993.
The VLBI coordinates of the host galaxy are\citemethod{deller2017} 13:09:47.69398 $-$23:23:02.3195, with estimated uncertainties dominated by systematics of $\lesssim$1 mas in each coordinate. 
We measured the HST centroid position of NGC 4993, but this was not trivial.
It is not clear what fraction of the central flux is in a point source and what fraction is in the background galaxy or nuclear star cluster.  
This affects how undersampled the central pixels are.  
We tried two ways to fit a central source in each of the four F160W exposures: 1) a simple centroid found using the very centermost set of pixels, and 2) to look for a point of symmetry in the annulus of pixels between radius$=$2 and radius$=$6.5 (in units of HST pixels).
The error bars come from the agreement among the four independent measurements (one for each exposure).
In the GAIA pixelized frame we measured these positions to be (X,Y) $\simeq$ (2500.5, 2500.5) and (2500.4, 2500.4) respectively with the uncertainty of $\lesssim$0.1 pixel in each axis. 
These positions are $\gtrsim$10 mas offset from the VLBI position $\simeq$ (2500.1, 2500.2) of NGC 4993, and we conclude that this discrepancy is due the inability to measure an accurate position for the nucleus (for reasons mentioned above) and/or due to a genuine offset between the positions of the optical nucleus and the radio core\citemethod{kovalev2017,petrov2019}.
The measurement of the NGC 4993's nucleus therefore does not provide any useful verification of the VLBI versus GAIA-CRF2 coordinates. 

\section{Parameter constraints from the point-source model}\label{sec:point-source}

We consider the motion of a certain part of a structured jet and ignore lateral expansion which can only be accurately captured by relativistic hydrodynamic simulations\citemethod{granot01_jet_dynamics, kumar03_jet_dynamics, zhang09_jet_simulation, vaneerten10_2Djet_simulation, decolle12_jet_simulation, duffell18_lateral_expansion, lu20_jet_dynamics, fernandez22_lateral_expansion}. At the time of radio astrometric measurements, the emitting material has already decelerated significantly from its initial Lorentz factor. This is because otherwise the flux contribution from the emitting material should rise rapidly with observer's time as $t^3$ (much steeper than the observed lightcurve), since the number of emitting electrons increases as $t^3$ for a circum-merger medium of constant density. Thus, the dynamics is given by the Blandford-McKee solution\citemethod{blandford76_dynamics}
\begin{equation}
  \label{eq:1}
  \Gamma \propto r^{-3/2},
\end{equation}
where $\Gamma$ is the Lorentz factor of the emitting gas, $r$ is the distance to the source, and we have assumed a constant density circum-merger medium. The relationship between the shock radius $r$ and lab-frame (or the rest frame of the compact object) time $t_{\rm lab}$ is
\begin{equation}
  \label{eq:2}
  r\approx c\int_0^{t_{\rm lab}}\left(1-{1\over 2\Gamma^2}\right) \d t_{\rm lab} \approx ct_{\rm lab}
  \left( 1 - {1\over 8\Gamma^2}\right),
\end{equation}
where we have used the approximated speed (in units of the speed of light $c$) $\beta\approx 1-1/(2\Gamma^2)$ and $\Gamma(t_{\rm lab})\propto r^{-3/2}\propto t_{\rm lab}^{-3/2}$ to the lowest order (affording an error of $\mathcal{O}(\Gamma^{-4})\sim 1\%$ or less). Suppose the angle between the velocity vector of the emitting material and the line of sight (LOS) is $\theta$, the observer's time is given by
\begin{equation}
  \label{eq:3}
  ct = ct_{\rm lab} - r\cos\theta \approx {r\over
    2\Gamma^2}\left(\Gamma^2\theta^2 + {1\over 4}\right).
\end{equation}
and the transverse separation between the flux centroid and the center of explosion is $r\sin\theta\approx r\theta$. Here we have made use of the approximations $\sin\theta\approx \theta$ and $1-\cos\theta\approx \theta^2/2$ with a fractional error of $\theta^2/6\sim \rm \rm 3\%$ or less, since $\theta< 24^{\rm o}$ as we will show later. The mean apparent speed since the explosion is given by
\begin{equation}
  \label{eq:5}
  \bar{\beta}_{\rm app} = {r\sin\theta\over ct} \approx
  {2\over \theta } \left(1 + {1\over 4\Gamma^2\theta^2}\right)^{-1}.
\end{equation}
Note that, if the velocity history $\beta(t_{\rm lab})$ is unknown, then the measured mean apparent speed since the explosion $\bar{\beta}_{\rm app}=t^{-1}\int_0^t \beta_{\rm app}\,\d t$ (time-averaging in the observer's frame) only constrains the mean physical speed $\bar{\beta}=t_{\rm lab}^{-1}\int_0^{t_{\rm lab}} \beta\,\d t_{\rm lab}$ (time-averaging in the lab frame), according to $\bar{\beta}_{\rm app} = \bar{\beta}\sin\theta/(1 - \bar{\beta}\cos\theta)$. In the limit $\bar{\Gamma}\equiv (1-\bar{\beta}^2)^{-1/2}\gg 1$ and $\theta\ll 1$, one obtains a conservative constraint $\bar{\beta}_{\rm app}\approx (2/\theta) (1 + \bar{\Gamma}^{-2}\theta^{-2})< 2/\theta$, which means the viewing angle is less than $2/\bar{\beta}_{\rm app}$, independent of the velocity history.
The two VLBI astrometric measurements at $t = 75$ and $230\rm\, d$, combined with our HST position of the merger, constrain the viewing angle $\theta$ and Lorentz factor $\Gamma$ of the emitting material at each of the epochs. To pin down each of the quantities, another relation between $\theta$ and $\Gamma$ is needed.

Note that at different epochs, the flux is generally dominated by different portions of the jet. Before the peak of the afterglow lightcurve, the flux is dominated by the jet region\citemethod{beniamini2020} where $\Gamma\theta\sim 1$ (a crude estimate to be better quantified later), which means that $\theta \sim 1.6/\bapp\sim 13^{\rm o}$ at $t =
75\rm\, d$. In the following, we provide a simple model for the probability distribution of the product $x\equiv \Gamma\theta$, based on the standard synchrotron afterglow theory\citemethod{kumar15_GRB_review}.

The characteristic synchrotron frequency of electrons with Lorentz
factor $\gamma$ in the comoving frame of the emitting plasma scales as
\begin{equation}
  \label{eq:6}
  \nu \propto \mc{D} \gamma^2B,
\end{equation}
where $B\propto \Gamma$ is the magnetic field strength in the comoving
frame and the Doppler boosting factor is given by
\begin{equation}
  \label{eq:7}
  \mc{D} = {1\over \Gamma(1-\beta\cos\theta)}\approx
  {2\Gamma \over 1 + \Gamma^2 \theta^2}.
\end{equation}
Electrons are accelerated by the shock into a power-law Lorentz factor distribution $\d N/\d\gamma \propto r^3 \gamma_{\rm m}^{-1} (\gamma/\gamma_{\rm m})^{-p}$ for $\gamma>\gamma_{\rm m}$, where the minimum Lorentz factor scales as $\gamma_{\rm m}\propto \Gamma$ and $r^3$ accounts for the volume of the gas swept up by the shock. In the optically thin limit, the flux as contributed by a given angular portion of the jet scales as   
\begin{equation}
  \label{eq:4}
  F_\nu \propto B r^3 (\gamma/\gamma_{\rm m})^{1-p} \propto
  \mc{D}^{{p+5\over 2}} \Gamma^{{3p-5\over 2}}\nu^{{1-p\over 2}}.
\end{equation}
The observed spectrum of $F_{\nu}\propto \nu^{-0.58}$ gives $p = 2.16$ to high precision\citemethod{margutti18_afterglow_spectrum, troja2020, makhathini2020}. At a fixed observing frequency, one has
\begin{equation}
  \label{eq:8}
  F_\nu \propto x^{2p}(1+x^2)^{-{p+5\over 2}}, \ \ x\equiv
  \Gamma\theta.
\end{equation}
This can be approximately considered as the likelihood function for $x$, because the total flux at a given time (before or near the lightcurve peak) is dominated by the brightest region of the jet. Therefore, we can estimate the probability density distribution of $\mathrm{ln}\, x$ by taking a flat prior in logarithmic space,
\begin{equation}
  \label{eq:9}
  {\d P_0\over \d\,\mathrm{ln}\,x} \propto x^{2p}(1+x^2)^{-{p+5\over 2}}.
\end{equation}
We take the prior on the viewing angle to be $\d P_0/\d\theta\propto \sin\theta$, and then the likelihood for each pair of ($x, \, \theta$) as drawn from the above distributions is given by a Gaussian of mean $\mu_{\bapp}$ and standard deviation $\sigma_{\bapp}$ (the measured mean and $1\sigma$ error) for the corresponding mean apparent speed $\bapp(x,\, \theta)$, according to the Bayesian Theorem. From this, we draw the posterior distribution of ($x, \, \theta$) using the $\mathtt{emcee}$ Markov-Chain Monte Carlo method\citemethod{emcee}. Furthermore, since we are seeing the emission from the most energetic part of the near the jet axis at 230 d and the emitting material at 75 d should be closer to the LOS, so we include an additional, conservative constraint of $\theta_{230\d} - \theta_{75\d}>0$ in our simulation. 

This method is directly applied to the proper motion measurement at $t=75\rm\,d$. However, the epoch at $t_{\rm obs}=230\rm\,d$ is observed after the peak of the lightcurve and hence the most energetic part (the ``core'') of the jet has likely already decelerated to a Lorentz factor slightly smaller than $\theta^{-1}$. Based on the Blandford-McKee dynamical evolution $\Gamma\propto t^{-3/8}$, we scale $x= \Gamma\theta$ drawn from Eq. (\ref{eq:9}) by a factor of $(230/175)^{-3/8}=0.86$ to remove the bias due to the deceleration of the jet core since the lightcurve starts to decline at $t_{\rm c}=175\rm\, d$, although our results are not sensitive (to within $2\%$) to the small uncertainties ($\pm 10 \rm\, d$) of the exact time the lightcurve starts to decline\citemethod{makhathini2020}. 

From the marginalized distributions, we find $\Gamma_{75\d} = 5.8^{+4.2}_{-1.9}$, $\theta_{75\d} = 13.9^{+3.3}_{-2.5}$ degrees and $\Gamma_{230\d} = 4.1^{+2.6}_{-1.2}$, $\theta_{230\d} = 20.2^{+2.8}_{-2.8}$ degrees (hereafter the errors are at 1$\sigma$ confidence). Since the difference between $\theta_{230\d}$ and $\theta_{75\d}$ should in fact be more than the size of the jet core, which is about $5^{\rm o}$ based on the lightcurve modeling (see \S \ref{sec:hydro}). This motivates us to try a more stringent prior of $\theta_{230\d} - \theta_{75\d}>5^{\rm o}$, and we find the final constraints on the inferred parameters are largely unchanged within the uncertainties. The results based on the more stringent prior, $\Gamma_{75\d} = 5.6^{+3.8}_{-1.7}$, $\theta_{75\d} = 12.8^{+2.5}_{-2.5}$ degrees and $\Gamma_{230d} = 4.7^{+3.1}_{-1.4}$, $\theta_{230\d} = 21.3^{+2.5}_{-2.3}$ degrees, are quoted in the Main Text. 


We show these constraints based on the prior of $\theta_{230\d} - \theta_{75\d}>5^{\rm o}$ in the $(\Gamma, \theta)$ plane for the two epochs in Figure~\ref{fig:speed_theta} and the schematic picture in Figure~\ref{fig:geometry}. 
The parameter values derived using the different priors are tabulated in Extended Data Table~\ref{tab:parameters}.
Finally, we combine the results from these different priors to obtain a robust constraint on the viewing angle (i.e. the angle between the Earth line of sight and the jet axis, or equivalently the inclination angle of the merger), $\theta_v=\theta_{230\d} \in (19^{\rm o},\ 24^{\rm o})$ at 1$\sigma$ confidence. We also applied the above analysis to the 206 d epoch data $\bar{\beta}_{\rm app}=4.7\pm 0.6$, which has larger fractional errors, and obtained a looser constraint $\theta_{\rm 206d} = 22.8^{+4.3}_{-3.8}$ deg, which is consistent with the viewing angle inferred from the 230 d data.

Note that the angle $\theta_{230\d}$ is the viewing angle, because we are directly measuring the position of the jet core at this epoch; whereas in the earlier epoch $t=75\,\d$, the emission comes from the less energetic \textit{wing} of the jet, which is $6^{\rm o}\mbox{--}11^{\rm o}$ ($1\sigma$, median $\approx 8^{\rm o}$) away from the jet axis. Since the emitting material at $t=75\,\d$ has already decelerated substantially from its original Lorentz factor, we see that the jet wing is initially highly relativistic with Lorentz factor $\Gamma_{i,75\d}>\bapp(75\d)\simeq 7$. Furthermore, the Lorentz factor of the jet core is even higher $\Gamma_{i,c}>10\mbox{--}20$, since its emission is strongly beamed away from us until much later (near the peak of the afterglow lightcurve). Our improved constraint on the inclination angle of GW170817, $\theta_{v}\in (19^{\rm o}, 24^{\rm o})$, rules out a substantial fraction of the parameter space allowed by the radio VLBI data alone.

Finally, since the Lorentz factor of the emitting material is directly constrained by our proper motion measurements, this allows us to robustly constrain the ratio between the isotropic equivalent energy for the jet core $E_{\rm iso}$ and the density of the pre-shock medium $n$ according to
\begin{equation}
  \label{eq:11}
  {E_{\rm iso}\over n_0} = {32\pi \over 3}m_p c^2 (ct)^3 \Gamma^8 \left(\Gamma^2\theta^2 + 1/4\right)^{-3},
\end{equation}
and from our marginalized posterior for $(\Gamma_{\rm 230d}, \theta_{\rm 230d})$, we obtain ${E_{\rm iso}/n_0} = 10^{55.8\pm 0.5}\rm\, erg\,cm^3$.


\section{Hydrodynamical Simulations}\label{sec:hydro}
We used the relativistic hydrodynamic code $\mathtt{Jedi}$\cite{lu20_jet_dynamics} to carry out about a million independent simulations of an axisymmetric, structured jet interacting with the circum-stellar medium, including the effects of lateral expansion. 

The advantage our hydrodynamic method over the semi-analytic point-source model in \S\ref{sec:point-source} is that it has the full jet angular structure under axisymmetry. This allows us to directly constrain the jet angular structure (although within our power-law jet parameterization, see below) by fitting to the full set of observational data, which is not possible for the semi-analytic model. Although the jet lateral expansion is intrinsically a 2D problem, the fact that the forward shock-compressed region is very thin in the radial direction motivates an effective 1D solution \cite{kumar03_jet_dynamics}. This approach is taken by the $\mathtt{Jedi}$ code, which is is much faster than other 2D codes in that each simulation only takes a few seconds on a CPU core --- this makes it possible to run $>10^6$ simulations to fit the data in a Monte Carlo manner.

The general jet structure has two functional degrees of freedom --- the angular structures of the kinetic energy and Lorentz factor. Afterglow data from GW170817, although extensive, does not provide sufficient information to inverse-reconstruct the full functional forms of the jet structure\cite{takahashi19}. 
Instead, we consider a power-law model which describes the full jet structure with 5 parameters: (as previously considered by Refs\cite{zhang02,rossi02,kumar03_jet_dynamics} and motivated by recent simulations by Ref\cite{gottlieb21_jet_structure})
\begin{equation}
    {\d E \over \d \Omega}(\theta) = {E_{\rm iso}\over 4\pi} \left[1 + \left({\theta/ \theta_{\rm c}}\right)^{2} \right]^{-q/2},
\end{equation}
\begin{equation}
    u_0(\theta) = u_{\rm 0,max} \left[1 + \left({\theta/ \theta_{\rm c}}\right)^{2} \right]^{-s/2},
\end{equation}
where $\thec$ is the half opening angle of the jet core (where most of the energy is contained), $\Eiso$ is the isotropic equivalent energy on the jet axis, $u_{\rm 0, max}$ is the maximum four-velocity on the jet axis, $q$ and $s$ are power-law indices describing how energy is distributed in the jet wing at $\theta\gg \thec$. The jet core Lorentz factor, as defined in the main text, is given by $\Gamma_{i,c}\approx u_{\rm 0, max}$ in the ultra-relativistic limit.

We adopt a constant circum-stellar medium (CSM) density $n_0$, as expected for old isolated double neutron star systems\cite{ramirez-ruiz19}. The other parameters include the observer's viewing angle $\theLOS$ with respect to the jet axis, luminosity distance to the source $D_{\rm L}$, the fractions of thermal energy in the shocked CSM that are shared by magnetic fields and shock-accelerated electrons $\epse$ and $\epsB$, and the power-law index $p$ for the Lorentz factor distribution of relativistic electrons. We fix $\epse=0.1$ as constrained by many previous works on GRB afterglow modeling\cite{panaitescu02}, so the entire model has 10 free parameters. However, since the entire spectrum from radio to the X-ray band is consistent with a single power-law without a statistically significant indication of the synchrotron cooling frequency, it is not possible to break the well-known degeneracy\cite{nakar02} between $\Eiso$, $n_0$ and $\epsB$ --- the observables only dependent on the combined quantity $\Eiso/[n_0\epsB^{(p+1)/(p+5)}]$. 
This is because the radius position of the forward shock $r$ and the Lorentz factor of the emitting gas $\Gamma$ at a given time only depend on the ratio of $\Eiso/n_0$, and the flux density at a given time and frequency depends on the number of shock-accelerated electrons (which depends on $r$ and $n_0$) and the power per unit frequency per electron radiating in the observer's band (which depends on $\Gamma$, $n_0$ and $\epsB$ through the magnetic field strength in the shock-heated region). Based on these considerations, we fix $n_0=10^{-2.5}\rm\, cm^{-3}$ and consider the ratio $\Eiso/n_0$ to be a single parameter --- this reduced the number of dimensions to 9. We have verified (by running additional simulations) that the choice of $n_0$ does not affect the constraints on the shape of the jet angular structure ($u_{\rm 0,max}, q, s$), energy-to-density ratio $\Eiso/n_0$, electron power-law index $p$, viewing angle ($\theLOS$), and the luminosity distance $D_{\rm L}$, within the errors. 
However, the magnetic equipartition parameter $\epsB$ cannot be fully constrained due to degeneracy, and the peak value of its posterior scales with our choice of $n_0$ as $\epsB\propto n_0^{-(p+5)/(p+1)\approx -2.7}$ (as the electron power-law index is well constrained to be $p=2.16\pm 0.01$).

For each set of parameters, we ran a full relativistic hydrodynamic simulation with the code $\mathtt{Jedi}$\cite{lu20_jet_dynamics}, which includes the effects of lateral expansion. 
Synchrotron emission, including the effects of self-absorption and synchrotron cooling, are calculated in a post-processing manner, which yields the lightcurve at arbitrary frequencies and the projected positions of the flux centroid at a given frequency at any observer's time. 
The results are then compared with the full lightcurve dataset of GW170817 collected by Ref\cite{makhathini2020} (version 04-May-2021 available on on the web\citemethod{gw170817_afterglow_data_url}) as well as the proper motion data obtained in this work. 
Each $3\sigma$ flux upper limit $F_{3\sigma}$ is approximated treated as a ``detection'' with zero mean flux and standard deviation of $F_{\rm 3\sigma}/3$. 
As for the proper motion data, we consider two independent time intervals of 75--230d (between two HSA epochs) and 0--230d (between HST and the last HSA epochs), which yields angular separations of $2.7\pm 0.3\rm\, mas$ and $5.07\pm 0.4\rm\, mas$ ($1\sigma$ errors), and we approximate the error distributions of these two measurements as Gaussian. 
For the purpose of minimizing the systematic error, when computing the proper motion, we use angular diameter distance $D_{\rm A} = D_{\rm L}/(1+z)^2$ with a cosmological redshift factor $z\approx 0.01$.

We took logarithmic flat priors on $\log u_{\rm 0,max}$, $\log \thec$, $\log \epsB$, $\log (\Eiso/n_0)$ and flat priors on $q$, $s$, $p$, $\cos\theLOS$. The luminosity distance of the host galaxy NGC 4993 has been constrained by Ref\cite{cantiello2018}, based on which we take the prior on $D_{\rm L}$ to be a Gaussian with mean $40.7\rm\,Mpc$ and variance $2.4\rm\, Mpc$. The prior boundaries are chosen to be sufficiently wide based on trial runs such that the marginalized posterior of each of the parameters is practically unaffected by our choice. An exception is the peak Lorenz factor $u_{\rm 0,max}$, which is limited to be less than $10^4$, although the upper limit of this parameter is unconstrained by the current data, since we only see the jet after it has already decelerated to a Lorentz factor of less than about 10. For this reason, the posteriors of most parameters are unaffected by our choice of upper boundary for $u_{\rm 0,max}$. The posterior (especially the $90\%$ lower limit) of the peak Lorentz factor may be affected by our choice of the $\log u_{\rm 0,max}$ prior as well as by the power-law form of the jet angular structure. However, we emphasize that the measurement of the mean apparent speed $\bar{\beta}_{\rm app, 0-75d}\simeq 7$ strongly argues for the jet core Lorentz factor to be $u_{\rm 0,max}\gg 7$, because: (1) to avoid fine-tuning, the material dominating the emission at $t=75\rm\, d$ must have been decelerating at time much earlier than 75 d, meaning that its initial Lorentz factor is greater than 7, and (2) the rising afterglow lightcurve before the peak time indicates that the observer is seeing progressively inner regions of the jet which has higher Lorentz factors (or narrower beaming angles) at smaller polar angles.

Then, our posteriors are sampled using the Dynamically Nested Sampling method provided by $\mathtt{dynesty}$\cite{speagle20_dynesty}, according to the $\chi^2$ residual obtained from the fit to all flux density and proper motion data (each data point carrying an equal weight). The full posterior is shown in Extended Data Figure~\ref{fig:mcmc}. 
The jet inclination angle is constrained to be $\theLOS = 21.9^{+3.3}_{-2.9}$ degrees (90\% credible interval), and the ratio between the on-axis isotropic equivalent jet energy and the CSM density is constrained to be $\Eiso/n_0 = 10^{56.1\pm 0.5}\rm\, erg\, cm^3$ (90\% credible), both in agreement with the results from our semi-analytic point-source model in the previous section.
The peak Lorentz factor of the jet is constrained to be $1.6<\log u_{\rm 0,max}<3.9$ (90\% credible). The upper limit is subjected to our prior of $\log u_{\rm 0,max}<4$, whereas the lower limit is physically constrained by the data (mainly proper motion measurements), as can be seen from the rapid drop of the probability distribution below $\log u_{\rm 0,max}\simeq 1.6$. Thus, we consider $u_{\rm 0,max}>40$ to be a robust lower limit that is not affected by our prior choice. The choices of different jet angular structures other than the power-law forms considered in this work may weakly affect this lower limit and this needs to be studied by future works.


We also note that the power-law index $s$ for the Lorentz factor structure of the jet wind is correlated with the peak Lorentz factor $u_{\rm 0,max}$, which is in agreement with the prediction by Ref\cite{beniamini2020}, in their equation (17). 

We further combine 
our modeling with gravitational wave data\citemethod{A17:H0,hotokezaka2019-h0} to obtain a revised standard-siren constraint on the Hubble constant $H_0$.
This parameter is related to the luminosity distance $D_{\rm L}$ and the recessional speed of the local Hubble flow $v_{\rm H}$ by
\begin{equation}
    D_{\rm L} \approx {v_{\rm H}\over H_0},
\end{equation}
where we have ignored higher order terms in the limit $z\ll 1$. We use the same Gaussian PDF for the Hubble flow speed as adopted by Refs.\citemethod{A17:H0, hotokezaka2019-h0} with mean $\lara{v_{\rm H}} = 3017\rm\, km\,s^{-1}$ and standard deviation $\sigma_{v_{\rm H}}= 166\rm\, km\, s^{-1}$, which come from the center of mass speed of NGC 4993 relative to the CMB frame $3327\pm72\rm\, km\, s^{-1}$ and peculiar velocity $-310\pm 150\rm\, km\, s^{-1}$. Thus, the final cumulative probability distribution of the Hubble constant is given by
\begin{equation}
    P(>H_0) = \int { \d v_{\rm H} \over \sqrt{2\pi \sigma_{v_{\rm H}}^2}} \mr{e}^{-{1\over 2} \left(\lara{v_{\rm H}} - v_{\rm H}\over \sigma_{v_{\rm H}}\right)^2} \int_{v_{\rm H}\over H_0} {\d P\over \d D_{\rm L}} \d D_{\rm L}.
\end{equation}
We obtain $H_0 = 71.5\pm 4.6\rm\, km\, s^{-1}$ based on this analysis. Our results are consistent with that from Ref\citemethod{hotokezaka2019-h0}, which is based on similar methods, but in this work we include the complete observational dataset and extensive hydrodynamic modeling.



\end{methods}


\bibliographystylemethod{naturemag}
\bibliographymethod{refsmethod}

\clearpage
\begin{addendum}
\item[Acknowledgements] 
The authors are grateful to Adam Deller for pointing out the required correction for radio VLBI positions, for careful reading of the manuscript, and for providing detailed comments.
The authors thank Alberto Krone-Martins for helpful discussions, Dale Frail for commenting on an early version of this manuscript, and Yamini Mooley for help with manuscript submission.
KPM is indebted to Gaura Nitay for providing the impetus to execute this project.
This research is based on observations made with the NASA/ESA Hubble Space Telescope obtained from the Space Telescope Science Institute, which is operated by the Association of Universities for Research in Astronomy, Inc., under NASA contract NAS 5–26555. 
These observations are associated with HST programs GO-14771, GO-14804, and GO-15329.
KPM was a Jansky Fellow of the National Radio Astronomy Observatory and his work is currently supported through the National Research Foundation Grant AST-1911199. 
WL was supported by the David and Ellen Lee Fellowship at Caltech and Lyman Spitzer, Jr. Fellowship at Princeton University.
\end{addendum}


\begin{addendum}
\item[Author Contributions] JA led the HST analysis. WL set up the semi-analytical and hydrodynamical models. KPM led the scientific analysis and interpretation. 
All authors discussed and wrote the paper.

 \item[Competing Interests] The authors declare that they have no
competing financial interests.

 \item[Correspondence] Correspondence and requests for materials should be addressed to K.P.M. (email: kunal@astro.caltech.edu) and J.A. (email: jayander@stsci.edu).
 
 \item[Reprints and Permissions] Reprints and permissions information is available at www.nature.com/reprints.
\end{addendum}

\clearpage

\begin{addendum}

\item[Data Availability] All HST data used in this work are available via MAST (https://mast.stsci.edu/). The minimum dataset consists of archival HST data from programs GO-14771, GO-14804, and GO-15329.

\item[Code Availability] The astrometric, semi-analytical point-source and hydrodynamical codes are currently being prepared for public release and are available from the corresponding authors upon request.
See \url{http://www.tauceti.caltech.edu/kunal/gw170817/} for updates.
\end{addendum}


\newpage

\setcounter{figure}{0}
\renewcommand{\figurename}{Extended Data Figure}
\renewcommand{\tablename}{Extended Data Table}

\begin{table*}
\small
\centering
\caption{{\bf Log of archival HST data used in this work}}
\label{tab:obs}
\begin{tabular}{llllllll}
\hline\hline
(1)         & (2)     & (3)   & (4)        & (5)     & (6)        & (7) & (8)\\
UT Date     &   T     &  Exp. & Instrument & Filter   & F$_\nu$  &SNR & Comments\\
            &   (d)   & (s)   &         &          & ($\mu$Jy)\\
\hline
2017 Aug 22.4 &     4.9   &   100$\times$3 & WFC3/IR &  F160W    &  216  &  372 & KN \\
2017 Aug 27.3 &   9.8   &  253$\times$4 & WFC3/IR    &  F160W   &  40   & 263 & KN  \\
\hline
2017 Dec 06.0  &   110   &   2264 & WFC3/UVIS &  F606W   &  0.11  &   & AG  \\
2018 Jan 01.6  &   137   &   2120 & ACS/WFC   &  F606W   &  0.08  &   & AG \\
2018 Jan 29.7 &   165   &   2372 & WFC3/UVIS &  F606W   &  0.09  &    8 & AG \\
2018 Feb 05.7  &   172   &   2400 & WFC3/UVIS &  F606W   &  0.08  &   & AG \\
2018 Mar 14.6 &   209   &   2432 & WFC3/UVIS &  F606W   &  0.08  &   & AG \\
\hline
\multicolumn{8}{p{5in}}{Columns: (1) Observation date (UT), (2) time post-merger in days, (3) total exposure time or single exposure time $\times$ number of exposures, (4) HST instrument, (5) HST filter, (6) flux density of GW170817, taken from refs\citemethod{cowperthwaite2017,tanvir2017,lamb2019,piro2019,fong2019} (this column is just for reference and is irrelevant to any of the analysis presented in this work), (7) signal-to-noise-ratio in a single exposure (for the AG data the SNR for the coadd F606W image, comprising of five epochs, is given), and (8) comments (KN=kilonova, AG=afterglow).}
\end{tabular}
\end{table*}

\clearpage
\begin{table*}
\scriptsize
\centering
\caption{{\bf Gaia DR2/EDR3 reference stars used for the F160W analysis}}
\label{tab:f160w_ref_stars}
\begin{tabular}{lllllllllll}
\hline\hline
S\# & Source ID & RA & $\sigma_{\rm RA}$ & Dec & $\sigma_{\rm Dec}$ & X$_{\rm GAIA}$   & $\sigma_{\rm X}$ & Y$_{\rm GAIA}$ & $\sigma_{\rm Y}$     & G\\
  &  & (deg) & (mas) & (deg) & (mas) & (pix) & (pix) & (pix) & (pix) &(mag)\\
\hline
1 & 3504021408852807040 & 197.4553796 & 0.19 & -23.3761553 & 0.13 & 2079.8512 & 0.0114 & 2980.9269 & 0.0071 & 18.70 \\
2 & 3504021378788617472 & 197.4413991 & 0.09 & -23.3837331 & 0.06 & 3234.3337 & 0.0052 & 2298.9503 & 0.0033 & 17.29 \\
3 & 3504021408852806784 & 197.4590053 & 0.09 & -23.3848506 & 0.06 & 1780.1554 & 0.0057 & 2198.2718 & 0.0035 & 17.64 \\
4 & 3504021378787675008 & 197.4534169 & 0.16 & -23.3926106 & 0.11 & 2241.4931 & 0.0101 & 1499.8207 & 0.0062  & 18.45 \\
5 & 3504021443212545536 & 197.4347542 & 0.15 & -23.3803979 & 0.10 & 3783.7930 & 0.0091 & 2599.1470 & 0.0058 & 18.34 \\
6 & 3504021172630185728 & 197.4517754 & 0.05 & -23.3972573 & 0.03 & 2377.0871 & 0.0029 & 1081.8298 & 0.0019 & 16.30 \\
\hline
\end{tabular}
\end{table*}

\clearpage
\begin{longtable}{rrrrrrrr}
\caption{{\bf Positional measurements and transformed positions for F160W}}
\label{tab:transformation}\\
\multicolumn{8}{p{4in}}{Columns: (1) Exposure number (Exp. 1--3 are from 22 August and 4--7 are from 27 August 2017), (2) Reference star number (see Extended Data Table~\ref{tab:f160w_ref_stars}; GW is GW170817), (3), (4) X and Y positions in raw HST image, (5), (6) X and Y positions in the HST distortion-corrected frame, (7), (8) X and Y positions transformed into the pixelized GAIA frame.}
\endlastfoot

\hline
(1) & (2) & (3) & (4) & (5) & (6) & (7) & (8) \\
N\# & S\# & X$_{\rm RAW}$ & Y$_{\rm RAW}$ & X$_{\rm COR}$ & Y$_{\rm COR}$ & X$^\prime_{\rm GAIA}$ & Y$^\prime_{\rm GAIA}$ \\
 & & (pix) & (pix) & (pix) & (pix) & (pix) & (pix) \\
\hline
1 & 1 & 345.766 & 686.738 & 326.364 & 687.952 & 2079.879 & 2980.948\\
1 & 2 & 716.591 & 532.853 & 741.823 & 533.240 & 3234.318 & 2298.937\\
1 & 3 & 299.035 & 415.071 & 274.423 & 415.664 & 1780.128 & 2198.252\\
1 & 4 & 470.333 & 212.546 & 465.253 & 215.197 & 2241.519 & 1499.837\\
1 & 5 & 859.150 & 661.403 & 903.137 & 663.019 & 3783.783 & 2599.143\\
1 & 6 & 532.462 & 81.517 & 533.755 & 87.030 & 2377.085 & 1081.829\\
1 & GW & 492.703 & 555.830 & 491.152 & 555.958 & 2500.202 & 2500.259\\
2 & 1 & 349.075 & 690.001 & 330.082 & 691.243 & 2079.861 & 2980.924\\
2 & 2 & 719.882 & 536.139 & 745.538 & 536.542 & 3234.323 & 2298.965\\
2 & 3 & 302.371 & 418.405 & 278.142 & 418.971 & 1780.139 & 2198.278\\
2 & 4 & 473.677 & 215.904 & 468.978 & 218.495 & 2241.518 & 1499.816\\
2 & 5 & 862.436 & 664.658 & 906.870 & 666.320 & 3783.788 & 2599.133\\
2 & 6 & 535.806 & 84.894 & 537.473 & 90.321 & 2377.085 & 1081.830\\
2 & GW & 496.008 & 559.125 & 494.866 & 559.263 & 2500.185 & 2500.228\\
3 & 1 & 352.421 & 693.313 & 333.842 & 694.585 & 2079.872 & 2980.930\\
3 & 2 & 723.211 & 539.470 & 749.297 & 539.889 & 3234.346 & 2298.945\\
3 & 3 & 305.734 & 421.762 & 281.891 & 422.301 & 1780.151 & 2198.259\\
3 & 4 & 477.045 & 219.302 & 472.729 & 221.832 & 2241.484 & 1499.848\\
3 & 5 & 865.740 & 667.948 & 910.627 & 669.657 & 3783.772 & 2599.137\\
3 & 6 & 539.182 & 88.323 & 541.227 & 93.664 & 2377.088 & 1081.828\\
3 & GW & 499.366 & 562.451 & 498.640 & 562.600 & 2500.191 & 2500.262\\
4 & 1 & 410.729 & 687.618 & 399.396 & 688.729 & 2079.881 & 2980.953\\
4 & 2 & 782.538 & 535.674 & 815.696 & 536.289 & 3234.333 & 2298.956\\
4 & 3 & 365.742 & 415.734 & 348.923 & 416.157 & 1780.145 & 2198.272\\
4 & 4 & 538.374 & 214.107 & 540.869 & 216.735 & 2241.483 & 1499.790\\
4 & 5 & 924.266 & 664.963 & 976.327 & 666.945 & 3783.784 & 2599.143\\
4 & 6 & 601.332 & 83.404 & 610.042 & 88.931 & 2377.088 & 1081.833\\
4 & GW & 558.521 & 557.499 & 564.918 & 557.651 & 2500.181 & 2500.264\\
5 & 1 & 414.719 & 689.098 & 403.884 & 690.220 & 2079.899 & 2980.960\\
5 & 2 & 786.536 & 537.151 & 820.190 & 537.783 & 3234.348 & 2298.925\\
5 & 3 & 369.760 & 417.249 & 353.411 & 417.657 & 1780.151 & 2198.248\\
5 & 4 & 542.405 & 215.604 & 545.354 & 218.207 & 2241.452 & 1499.826\\
5 & 5 & 928.254 & 666.410 & 980.828 & 668.428 & 3783.773 & 2599.158\\
5 & 6 & 605.369 & 84.911 & 614.524 & 90.404 & 2377.091 & 1081.830\\
5 & GW & 562.519 & 558.971 & 569.406 & 559.131 & 2500.160 & 2500.224\\
6 & 1 & 413.228 & 691.585 & 402.211 & 692.736 & 2079.877 & 2980.920\\
6 & 2 & 785.024 & 539.661 & 818.524 & 540.292 & 3234.327 & 2298.939\\
6 & 3 & 368.271 & 419.767 & 351.747 & 420.164 & 1780.149 & 2198.268\\
6 & 4 & 540.885 & 218.171 & 543.676 & 220.728 & 2241.471 & 1499.833\\
6 & 5 & 926.722 & 668.909 & 979.144 & 670.943 & 3783.800 & 2599.157\\
6 & 6 & 603.856 & 87.491 & 612.863 & 92.916 & 2377.089 & 1081.829\\
6 & GW & 561.015 & 561.489 & 567.732 & 561.656 & 2500.138 & 2500.239\\
7 & 1 & 409.232 & 690.110 & 397.716 & 691.250 & 2079.903 & 2980.947\\
7 & 2 & 781.014 & 538.194 & 814.017 & 538.807 & 3234.332 & 2298.950\\
7 & 3 & 364.247 & 418.267 & 347.252 & 418.680 & 1780.141 & 2198.260\\
7 & 4 & 536.850 & 216.645 & 539.186 & 219.227 & 2241.463 & 1499.826\\
7 & 5 & 922.732 & 667.458 & 974.641 & 669.456 & 3783.784 & 2599.133\\
7 & 6 & 599.808 & 85.981 & 608.370 & 91.439 & 2377.091 & 1081.830\\
7 & GW & 557.005 & 560.007 & 563.231 & 560.166 & 2500.195 & 2500.229\\
\hline
\end{longtable}

\clearpage
\begin{table*}
\footnotesize
\centering
\caption{{\bf GW170817 positions and associated uncertainties at different epochs in the GAIA/ICRF3 reference frame}}
\label{tab:coordinates}
\begin{tabular}{llllll}
\hline
(1)   & (2) & (3)         & (4)         & (5)        & (6)\\       
Epoch & Telescope & Coordinates & Statistical & Systematic (corr.) & Systematic \\
\hline
8   & HST & 13:09:48.068473 $-$23:22:53.3906 & (0.32, 0.19) & \ldots & \ldots\\
75  & HSA & 13:09:48.068648 $-$23:22:53.3907 & (0.12, 0.4) & (0.18, 0.34) & (0.15, 0.5) \\
159  & HST & 13:09:48.06809\,\,\, $-$23:22:53.383 & (13, 11) & \ldots & \ldots\\
206 & gVLBI & 13:09:48.068770 $-$23:22:53.3906 & (0.21, 0.25) & (0.18, 0.34) & (0.15, 0.5) \\
230 & HSA & 13:09:48.068841 $-$23:22:53.3905 & (0.17, 0.4) & (0.18, 0.34) & (0.15, 0.5) \\
\hline
\multicolumn{6}{p{6in}}{Columns: (1) Mean observing epoch (days), (2) Telescope used for the measurement, (3) source coordinates in the GAIA or ICRF3 reference frames, (4) statistical measurement error on the source position, (5) systematic error, which is correlated between the three radio epochs (75 d, 206 d and 230 d), arising from the uncertainty in the position of the common phase reference source (J1312-2350, used to bring the radio positions of GW170817 to the ICRF3 frame), and (6) systematic error (uncorrelated) due to ionospheric contribution and phase referencing between J1321-2350 and GW170817. All uncertainties are given in the format: (RA mas, Dec mas).}
\end{tabular}
\end{table*}



\clearpage
\begin{table*}
\centering
\caption{{\bf GW170817 structured jet parameter values derived from the semi-analytical point-source model.}}
\label{tab:parameters}
\begin{tabular}{|c|c|c|}
\hline
Parameter & $\theta_{230\d} - \theta_{75\d}>0^{\rm o}$ prior & $\theta_{230\d} - \theta_{75\d}>5^{\rm o}$ prior\\
\hline
$\theta_{75\d}$ (deg) & $13.9_{-2.5}^{+3.3}$ & $12.8_{-2.5}^{+2.5}$\\
$\Gamma_{75\d}$ & $5.8_{-1.9}^{+4.2}$ & $5.6_{-1.7}^{+3.8}$\\
$\theta_{230\d}$ (deg) & $20.2_{-2.8}^{+2.8}$ & $21.3_{-2.3}^{+2.5}$\\
$\Gamma_{230\d}$ & $4.1_{-1.2}^{+2.6}$ & $4.7_{-1.4}^{+3.1}$\\
\hline
\end{tabular}
\end{table*}



\clearpage
\includegraphics[width=6.6in]{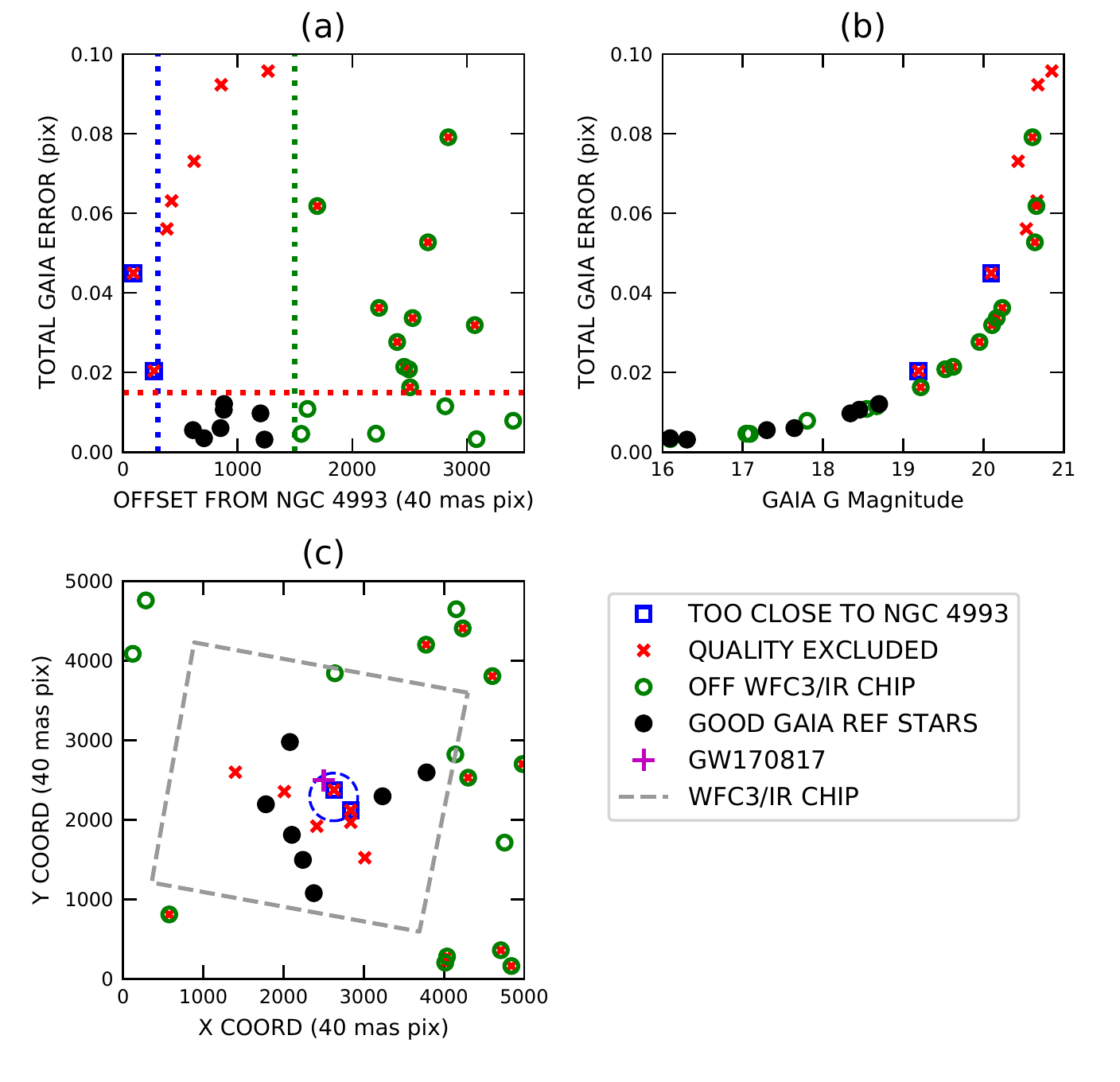}

\clearpage
\begin{figure}
\caption{{\bf Selection of GAIA reference stars for the F160W analysis.}
The panels (a), (b) give the positions, magnitudes and positional uncertainties (1$\sigma$) associated with the 32 GAIA stars that are within the WFC3/IR frame, which is shown in panel (c).
The legend shows the marker shape and color used for plotting these stars based on their vetted classifications.
The 6 GAIA reference stars selected based on low quoted GAIA positional errors, distant location from the host galaxy nucleus ($>$12 arcseconds from the nucleus of NGC 4993), centroid located on the HST chip, and away from any bad pixels, are shown as black filled circles.
In panels (a), (c) the blue dashed lines denote the 12 arcsecond distance constraint from the NGC 4993 nucleus, and the green dashed lines denote the extent of the WFC3/IR chip.}
\label{fig:gaia_sel}
\end{figure}



\clearpage
\centering
\includegraphics[width=6.6in]{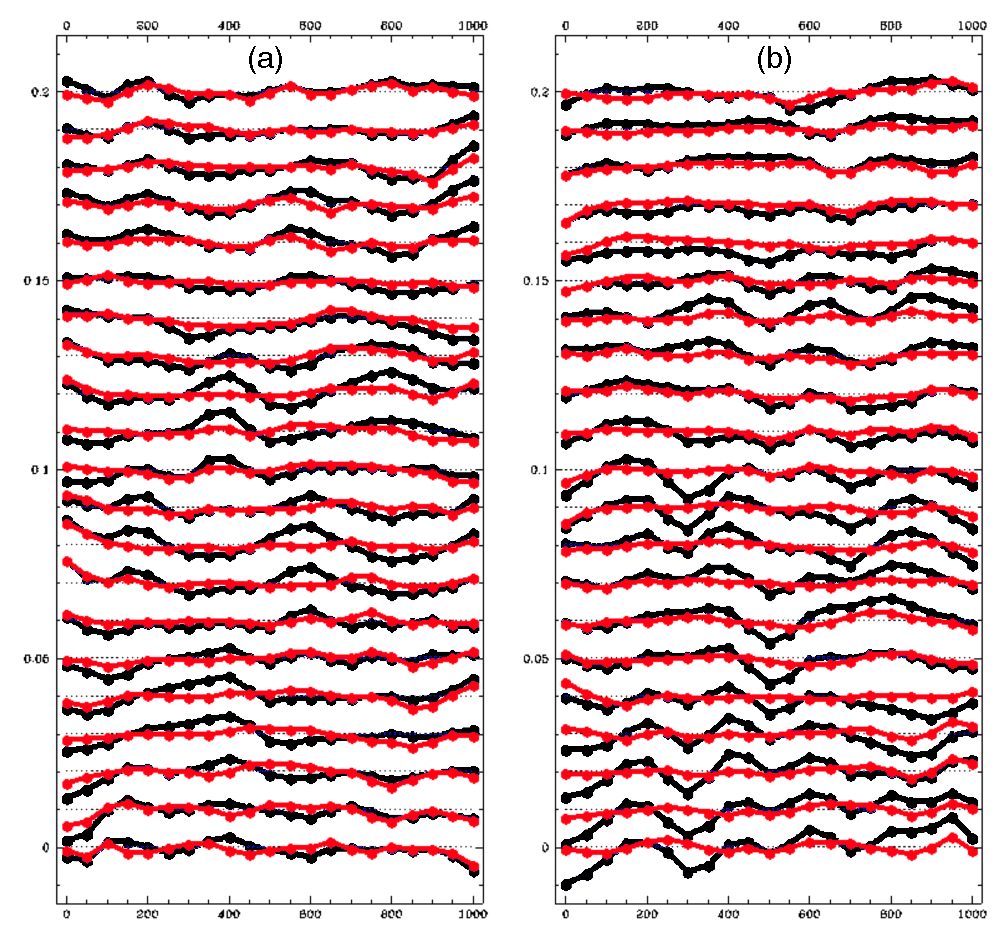}

\begin{figure}
\caption{{\bf Residuals from the distortion correction for WFC3/IR.}
The distortion residuals along each axis (image X/Y) for image slices that are 50-pixels wide in the orthogonal direction (see Methods for details).
The X residuals are shown in panel (a) and the Y residuals in panel (b).  
The horizontal axis in each panel represents the pixel number and the vertical axis represents the residual in units of pixels.
Each set of red and black curves, as well as each data point plotted on the red and black curves, represents one slice (offset of each set of curves along the vertical axis is arbitrary).
The black points/curves denote the distortion residuals after the standard HST distortion correction\cite{anderson2016} and the red after our improved correction.  
In general, the residuals went down by a factor of two in each coordinate after the application of the improved correction. 
The new distortion-correction residuals lie within 0.002 pixel per coordinate (i.e. within 0.08 mas; RMS).}
\label{fig:distortion_correction}
\end{figure}

\clearpage
\begin{figure}
\centering
\includegraphics[width=6.6in]{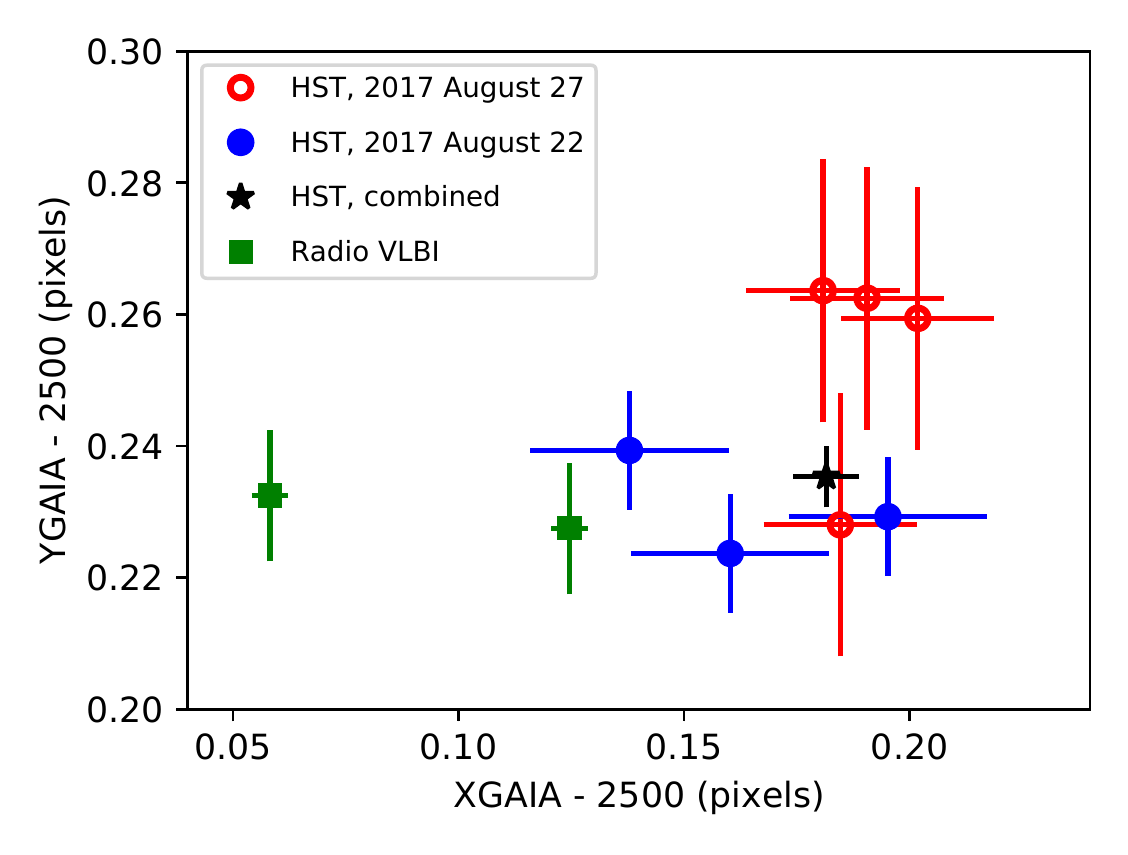}
\caption{{\bf HST/GAIA merger position of GW170817.}
The positions of GW170817 in the individual HST F160W exposures (blue filled and red unfilled circles; mean epoch 8 d post-merger) and the combined HST position (black star), in the GAIA pixelized frame, shown along with the radio VLBI measurements\cite{mooley2018-vlbi} at 75 d and 230 d.
The errorbars represent 1$\sigma$ statistical uncertainties.
The VLBI systematic uncertainties have not been included.}
\label{fig:consolidated_positions}
\end{figure}

\clearpage
\begin{figure}
\centering
\includegraphics[width=6.6in]{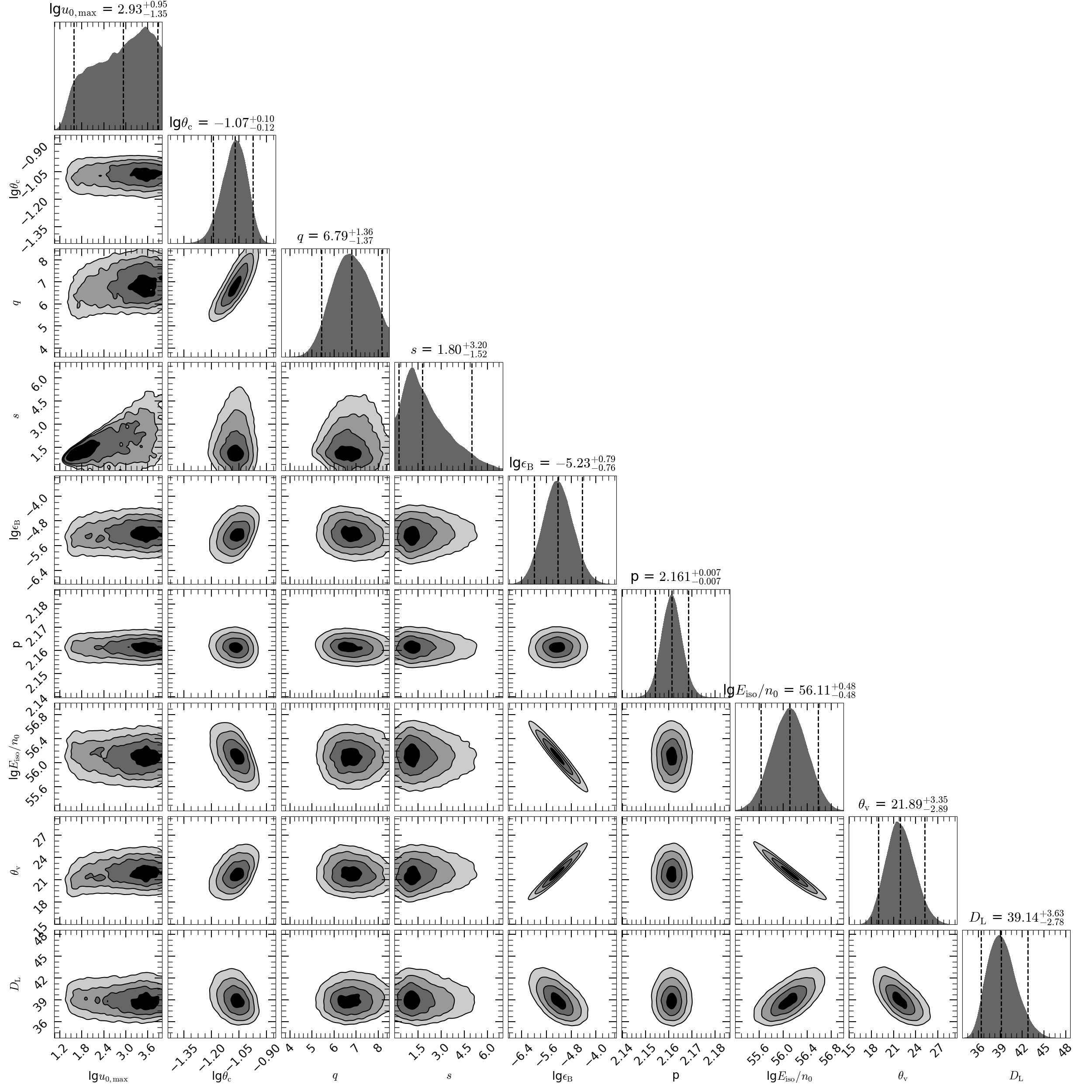}
\caption{{\bf Full posterior from the hydrodynamic simulations.} The parameters are: peak Lorentz factor $\mr{lg}u_{\rm 0,max}$, angular size of the jet core $\mr{lg}\thec\rm\,[rad]$, power-law index $q$ for the energy distribution of the jet wing,  power-law index $s$ for the Lorentz factor distribution of the jet wind, magnetic field equipartition parameter $\mr{lg}\epsB$, power-law index $p$ for the electron Lorentz factor distribution, $\mr{lg}\Eiso/n_0\rm\, [erg\,cm^3]$ --- ratio between the isotropic-equivalent energy on the jet axis and the CSM number density, inclination angle $\theLOS\rm\, [degree]$  between the line of sight and the jet axis, luminosity distance to the source $D_{\rm L}$. The dashed lines in the marginalized probability distributions indicate the 90\% credible interval for each parameter. }
\label{fig:mcmc}
\end{figure}

\end{document}